**Conformally Invariant**

**Scalar-Tensor Field Theories**

**in a**

**Four-Dimensional Space**

by

Gregory W. Horndeski
2814 Calle Dulcinea
Santa Fe, NM 87505-6425

e-mail:
horndeskimath@gmail.com

June13, 2017

# ABSTRACT

In a four-dimensional space , I shall construct all of the conformally invariant scalar-tensor field theories, which are flat space compatible; *i.e.,* well-defined and differentiable when evaluated for a flat metric tensor and constant scalar field. It will be shown that all such field theories must be at most of fourth-order in the derivatives of the field variables. The Lagrangian of any such field theory can be chosen to be a linear combination of four conformally invariant scalar-tensor Lagrangians, with the coefficients being functions of the scalar field. Three of these "generating" Lagrangians are of second-order, while one is of third-order. However, the third-order Lagrangian differs from a non-conformally invariant second-order Lagrangian by a divergence. Consequently, all of the conformally invariant, flat space compatible, scalar-tensor field theories, can be obtained from a second-order Lagrangian.

# TABLE OF CONTENTS

**Section 1: Introduction**

This paper will be concerned with scalar-tensor field theories in a four-dimensional space. The field variables will be the local components of the metric tensor, $g_{ab}$, and a real valued scalar field, $\varphi$. The equations of this theory will be assumed to be the Euler-Lagrange equations associated with a Lagrangian, which we take to be a scalar density concomitant of $g_{ab}$ and $\varphi$, along with their derivatives of arbitrary, albeit finite, order. The theory will be said to be of $k^{th}$ order, if there exist $k^{th}$ order derivatives of at least one of the field variables, in at least one of the sets of Euler-Lagrange tensor densities. The signature of the metric tensor will not be important in what we do, so I shall assume that it is arbitrary, but fixed. The notation that I use throughout is the same as that used in [1], unless stipulated to the contrary.

During the past few years there has been some work done on third-order scalar tensor field theories, which are often referred to as "Beyond Horndeski Theories." For some papers in this field, please see Zumalacárregui & Garcia-Bellido [2]; Gleyzes, *et al.,* [3], [5]; Lin, *et al.,* [4]; Deffayet, *et al.,* [6]; Chrisostomi, *et al.,* [7]; Bettoni & Zumalacárregui [8]; Tian & Booth [9]; Ezquiaga, *et al.,* [10], and Horndeski [11]. Many of the nine classes of Lagrangians that I presented in [11], appeared earlier in [10], while a special form of what I called the Pontrjagin scalar-tensor Lagrangian in [11], was studied by Tian & Booth in [9].

The purpose of this paper is to examine what effect the demand of conformal invariance has on the class of all scalar-tensor field theories. The conformal transformations we shall consider will be of the form: $g_{ab} \rightarrow g'_{ab} := e^{2\sigma} g_{ab}$, where $\sigma$ is a differentiable scalar field. Under such a transformation a Lagrangian of the form

$$L = L(g_{ab}; g_{ab,c}; \ldots; \varphi; \varphi_{,c}; \ldots)$$

transforms to a new Lagrangian L' defined by

$$L'(g'_{ab}; g'_{ab,c}; \ldots; \varphi; \varphi_{,c}; \ldots) := L(g'_{ab}; g'_{ab,c}; \ldots; \varphi; \varphi_{,c}; \ldots) \ .$$

L is said to be conformally invariant if $L' = L$. When L is conformally invariant, it is well-known that $E^a_b(L)$ and $E(L)$ are also conformally invariant. (Proof for $E^a_b(L)$:

$$(E^{ab}(L))' \equiv \frac{\delta L'}{\delta g'_{ab}} = \frac{\delta L}{\delta g'_{ab}} = \frac{\delta L}{\delta g_{rs}} \ \frac{\delta g_{rs}}{\delta g'_{ab}} = E^{ab}(L) \ e^{-2\sigma} \ .)$$ When $E^a_b(L)$ and $E(L)$ are conform-

ally invariant we shall say that a scalar-tensor field theory is conformally invariant. More will be said about the relationship between conformally invariant Lagrangians, and conformally invariant field theories later.

Let us now quickly review a few of the things I did in [11], since that will be pertinent to what follows. There a great deal of time was devoted to the study of the Pontrjagin scalar-tensor Lagrangian

$$L_P := p(\varphi, \rho) \ P \qquad\qquad\qquad\qquad Eq.1.1$$

where

$$P := \varepsilon^{abcd} \, R^{pq}{}_{ab} \, R_{pqcd} \qquad\qquad \text{Eq.1.2}$$

is the Pontrjagin Lagrangian [12] in a 4-dimensional space, $\varepsilon^{abcd}$ is the Levi-Civita tensor density, and $p = p(\varphi, \rho)$ is a differentiable function of $\varphi$ and $\rho := g^{ab}\varphi_{,a}\varphi_{,b}$. When p is independent of $\rho$, the Lagrangian $L_p$ becomes conformally invariant, since P can be rewritten as

$$P = \varepsilon^{abcd} \, C^{pq}{}_{ab} \, C_{pqcd} \qquad\qquad \text{Eq.1.3}$$

where $C_{hijk}$ is the Weyl tensor, defined in a four-dimensional space by

$$C_{hijk} := R_{hijk} + \tfrac{1}{2}(g_{hk} \, R_{ij} + g_{ij} \, R_{hk} - g_{hj} \, R_{ik} - g_{ik} \, R_{hj}) + \tfrac{1}{6} R \,(g_{hj}\, g_{ik} - g_{hk}\, g_{ij}). \quad \text{Eq.1.4}$$

Since $C_{h}{}^{i}{}_{jk}$ is conformally invariant, it should be obvious why

$$L_{3C} := p(\varphi) \, P = p(\varphi) \, \varepsilon^{abcd} \, C^{pq}{}_{ab} \, C_{pqcd} \qquad\qquad \text{Eq.1.5}$$

is conformally invariant. I shall refer to $L_{3C}$ as the Pontryjagin conformally invariant scalar-tensor Lagrangian. From [11] we find that

$$E^{ab}(L_{3C}) = \ 4p' \, \varphi_h \, [\varepsilon^{rsah} \, R_r{}^{b}{}_{|s} + \varepsilon^{rsbh} \, R_r{}^{a}{}_{|s}] +$$
$$\qquad - \ 2(p''\varphi_h \, \varphi_k + p'\varphi_{hk})[\varepsilon^{rsah} \, R_{rs}{}^{kb} + \varepsilon^{rsbh} \, R_{rs}{}^{ka}] \,, \qquad \text{Eq.1.6}$$

and

$$E(L_{3C}) = \ - \, p' \, P \qquad\qquad \text{Eq.1.7}$$

where " $'$ " denotes a derivative with respect to $\varphi$, and, for convenience, I have dropped the bars on covariant derivatives of $\varphi$, so $\varphi_{hk} := \varphi_{|hk}$. It is obvious that $E(L_{3C})$ is conformally invariant, but that is not obviously the case with $E^{a}{}_{b}(L_{3C})$. Fortunately

it must be so, since $L_{3C}$ is conformally invariant, and hence $L_{3C}$ generates a conformally invariant, third-order scalar-tensor field theory.

I should point out that I am using the Lovelock and Rund [13] conventions for the variational derivative, which is based on the Carathéodory approach to the calculus of variations. This explains the minus sign in Eq.1.7.

To compute the above Euler-Lagrange tensor densities I drew upon the formulas presented in [1], which were derived in detail in [14] (*see,* page 116). According to these formulas, if L is a second-order scalar-tensor Lagrangian, then its Euler-Lagrange tensor densities can be expressed in a manifestly tensorial manner as follows:

$$E^{ab}(L) = -\Pi^{ab,hk}{}_{|hk} + \Pi^{ab,h}{}_{|h} - \Pi^{ab} \qquad\qquad \text{Eq.1.8}$$

and

$$E(L) = -\zeta^{hk}{}_{|hk} + \zeta^{h}{}_{|h} - \zeta \qquad\qquad \text{Eq.1.9}$$

where

$$\Pi^{ab,hk} := \frac{\partial L}{\partial g_{ab,hk}} \; ; \; \zeta^{hk} := \frac{\partial L}{\partial \varphi_{,hk}} \; ; \; \zeta := \frac{\partial L}{\partial \varphi} , \qquad\qquad \text{Eq.1.10}$$

$$\Pi^{ab,h} = \tfrac{1}{2}(\zeta^{ab}\,\varphi^h - \zeta^{hb}\,\varphi^a - \zeta^{ah}\,\varphi^b) ; \qquad\qquad \text{Eq.1.11}$$

$$\Pi^{ab} = \tfrac{1}{3}R_k{}^b{}_{mh}\,\Pi^{hk,am} - R_k{}^a{}_{mh}\,\Pi^{hk,bm} -\tfrac{1}{2}\,\varphi^a\,\zeta^b - \zeta^{bk}\varphi_k{}^a + \tfrac{1}{2}g^{ab}\,L, \quad \text{Eq.1.12}$$

and

$$\zeta^a := \frac{\partial L}{\partial \varphi_{,a}} + \zeta^{hk} \, \Gamma^a_{\ hk} \, , \qquad\qquad \text{Eq.1.13}$$

with $\Gamma^a_{\ hk}$ denoting the components of the Christoffel symbols of the second kind. (Note that in Eq.4.11 of [1] there is a typographical error in the equation for $\Pi^{ab}$ involving $\varphi^a \zeta^b$, which is corrected in Eq.1.12 above.)

If you are not familiar with tensorial concomitants, and the process of differentiating them with respect to their various arguments, please see Appendix A, where this topic is discussed.

One of the interesting properties of $L_{3C}$ and its Euler-Lagrange tensor densities, is that they all vanish in a flat space, leaving no trace of the scalar field. Consequently, one can not find $L_{3C}$, $E^{ab}(L_{3C})$ or $E(L_{3C})$, by trying to reconstruct them from their flat space remnants, as was done by Deffayet, *et al.*, in [15], to construct second-order scalar-tensor field theories, from their flat space counterparts.

In [11] I apply a simple form of Beckenstein's [16] disformal transformation to the Lagrangian $L_P$, given in Eq.1.1, to produce another class of second-order Lagrangians that yield third-order Euler-Lagrange tensor densities. Amongst this class of Lagrangians is

$$L_{PC2} := f(\varphi) \, \varepsilon^{rstu} \, \varphi_u \, \varphi^p \, \varphi_{qt} R_p^{\ q}_{\ rs} \, / \, \rho \qquad\qquad \text{Eq.1.14}$$

where f is a differentiable scalar function of $\varphi$. $L_{PC2}$ is conformally invariant, however it, and its associated Euler-Lagrange tensor densities, are not well-defined for all choices of the scalar field.

If we examine the second-order scalar-tensor field theories I presented in [1], we would find that only one Lagrangian there yields a conformally invariant scalar-tensor field theory; *viz*., the conformally invariant Lagrangian

$$L_{2C} := g^{\frac{1}{2}} k(\varphi) \rho^2 , \qquad\qquad \text{Eq.1.15}$$

where k is a differentiable scalar function of $\varphi$. The Euler-Lagrange tensor densities associated with $L_{2C}$ are given by

$$E^{ab}(L_{2C}) = g^{\frac{1}{2}} k (2\rho \, \varphi^a \, \varphi^b \ - \tfrac{1}{2}\rho^2 \, g_{ab} ) \qquad\qquad \text{Eq.1.16}$$

and

$$E(L_{2C}) \quad = 4g^{\frac{1}{2}} k \, \rho \Box \varphi + 8g^{\frac{1}{2}} k \, \varphi_{ab} \, \varphi^a \, \varphi^b + 3g^{\frac{1}{2}} k' \, \rho^2 , \qquad\qquad \text{Eq.1.17}$$

which are of second-order. We see that $L_{2C}$, and its associated Euler-Lagrange tensor densities are well-defined and differentiable, for all values of the metric tensor and scalar field. Similarly, this is also the case for $L_{3C}$ , and its Euler-Lagrange tensor densities.

We arrived at $L_{3C}$ by starting with a conformally invariant metric Lagrangian. Another such Lagrangian is the one which generates the Bach tensor $B^{ab}$(*see*, [17]), which is defined by

$$g^{\frac{1}{2}} B^{ab} := E^{ab}(-\tfrac{1}{2} \, g^{\frac{1}{2}} \, C^{hijk} \, C_{hijk})$$

$$= g^{\frac{1}{2}} (C^{acdb}{}_{|cd} + C^{adcb}{}_{|cd} - R_k{}^a{}_{mh} \, C^{kbmh} + \tfrac{1}{4} \, g^{ab} \, C^{hijk} \, C_{hijk} ) .$$

Evidently, $g^{\frac{1}{2}} B^a{}_b$, is a fourth-order, conformally invariant tensor density. We now let

$$L_{4C} := g^{\frac{1}{2}} b(\varphi) C^{hijk} C_{hijk} , \qquad\qquad \text{Eq.1.18}$$

where b is a differentiable scalar function of $\varphi$. The scalar-tensor field theory that $L_{4C}$ generates, will be referred to as the Bach conformally invariant scalar-tensor field theory. Using Eqs.1.8-1.13 we find that

$$E^{ab}(L_{4C}) = -2\, g^{\frac{1}{2}}\, b\, B^{ab} - 4\, g^{\frac{1}{2}}\, b'(C^{adcb}{}_{|c}\, \varphi_d + C^{acdb}{}_{|c}\, \varphi_d + C^{acdb}\, \varphi_{cd}\,) +$$
$$- 4\, g^{\frac{1}{2}}\, b''\, C^{acdb}\, \varphi_c\, \varphi_d , \qquad\qquad \text{Eq.1.19}$$

and

$$E(L_{4C}) = -\, g^{\frac{1}{2}}\, b'\, C^{hijk}\, C_{hijk} . \qquad\qquad \text{Eq.1.20}$$

In passing I would like to point out that in a four-dimensional space, the Bach tensor can be rewritten as

$$B^{ab} = -\,\Box R^{ab} + \tfrac{1}{3} R^{|ab} + \tfrac{1}{6}\, g^{ab}\Box\, R + 2R^{habk}\, R_{hk} + \tfrac{1}{2} g^{ab} R^{cd} R_{cd} +$$
$$+ \tfrac{2}{3} R R^{ab} - \tfrac{1}{6} g^{ab}\, R^2 . \qquad\qquad \text{Eq.1.21}$$

Like the theories generated by $L_{2C}$ and $L_{3C}$, $L_{4C}$ is well-defined, and differentiable, for all choices of the metric tensor and scalar field.

Now we could continue to use $g^{\frac{1}{2}}$, $\rho$ and $I := C^{hijk} C_{hijk}$, to generate infinitely many conformally invariant scalar-tensor field theories. To see how, let

$$L_{\alpha,\beta} := f_{\alpha,\beta}(\varphi)\, g^{\frac{1}{2}}\, \rho^\alpha\, I^\beta ,$$

where $f_{\alpha,\beta}$ is a differentiable function of $\varphi$. If $\alpha$ and $\beta$ are real numbers chosen so that $4 = 2\alpha + 4\beta$, then $L_{\alpha,\beta}$ is conformally invariant. However, we note that the Lagrangians $L_{\alpha,\beta}$ are only well-behaved and differentiable when $(\alpha,\beta) = (2,0)$, or $(0,1)$, which give rise to the Lagrangians $L_{2C}$ and $L_{4C}$.

I will now define a scalar-tensor field theory to be flat space compatible, if the Euler-Lagrange tensor densities of that theory are well-defined and differentiable when evaluated for a flat metric tensor, and constant scalar field. $L_{2C}$, $L_{3C}$ and $L_{4C}$ generate conformally invariant, flat space compatible, scalar-tensor field theories.

When you think about it, most physical field theories are flat-space compatible. *E.g.,* Einstein's vacuum field equations with cosmological term, are well-defined and differentiable in empty space. Similarly, the Einstein-Maxwell, and the Einstein-Yang-Mills field equations are well-defined and differentiable, when evaluated for a flat metric tensor, and constant vector potentials. This is also true for my generalizations of the Einstein-Maxwell, and Einstein-Yang-Mills equations, presented in [18] and [19]. Lastly, the Brans-Dicke theory [20], provides us with a flat-space compatible, scalar-tensor field theory. I believe that most physicists would regard a physical field theory to be absurd, if its equations blew up, when nothing was going on in space.

If we turn our attention to trying to devise flat-space compatible, conformally invariant, scalar-tensor field theories, one is hard pressed to imagine any such theories other than those presented so far. While attempting to prove that was the case, I did encounter another Lagrangian that generated such a theory. The Lagrangian is given by

$$L_{UC} := -12 \, g^{\frac{1}{2}} \, u \, R^{hk} \, \varphi_h \, \varphi_k + 2 \, g^{\frac{1}{2}} \, u \, \rho \, R - 3 \, g^{\frac{1}{2}} \, u \, (\Box\varphi)^2 - 6 \, g^{\frac{1}{2}} \, u \, \varphi^{hk} \, \varphi_{hk} +$$

$$- 12 \, g^{\frac{1}{2}} \, u \, \varphi^h \, \varphi^k_{\ kh} \ , \qquad\qquad\qquad\qquad\qquad \text{Eq.1.22}$$

and its associated Euler-Lagrange tensor densities are

$$E^{ab}(L_{UC}) =$$

$$g^{\frac{1}{2}} \, u \, [\, 2 \, \varphi^{abc} \, \varphi_c + \varphi^c_{\ cd} \, \varphi^d \, g^{ab} - 3 \, \varphi^c_{\ c}{}^a \, \varphi^b - 3 \, \varphi^c_{\ c}{}^b \, \varphi^a \,] + g^{\frac{1}{2}} \, u \, [-6 \, \varphi^a \, R^{bc} \, \varphi_c - 6 \, \varphi^b \, R^{ac} \, \varphi_c +$$

$$- 4 \, R^{cadb} \, \varphi_c \, \varphi_d + 4 \, R^{cd} \, \varphi_c \, \varphi_d \, g^{ab} + 2 \, \rho \, R^{ab} + 2 \, R \, \varphi^a \, \varphi^b - \rho \, R \, g^{ab} \,] + g^{\frac{1}{2}} \, u \, [\, 6 \, \varphi^{ab} \, \Box\varphi +$$

$$- \tfrac{3}{2} \, g^{ab} \, (\Box\varphi)^2 + g^{ab} \, \varphi_{cd} \, \varphi^{cd} - 4 \, \varphi^a_{\ c} \, \varphi^{bc}\,] + g^{\frac{1}{2}} \, u'[\, 4 \, \rho \, \varphi^{ab} + 2 \, \varphi^{cd} \, \varphi_c \, \varphi_d \, g^{ab} - 4 \, \varphi^a \, \varphi^{bc} \, \varphi_c +$$

$$- 4 \, \varphi^b \, \varphi^{ac} \, \varphi_c \, - \rho \, g^{ab} \, \Box\varphi \,] + 2 \, g^{\frac{1}{2}} \, u'' \, [\, \rho^2 \, g^{ab} - 4 \, \rho \, \varphi^a \, \varphi^b \,] \qquad\qquad \text{Eq.1.23}$$

and

$$E(L_{UC}) =$$

$$g^{\frac{1}{2}} \, u \, [-2 \, R_{|a} \, \varphi^a - 12 \, R^{ab} \varphi_{ab} + 4 \, R \, \Box\varphi - 6 \, \Box\Box\varphi\,] + g^{\frac{1}{2}} \, u' \, [-12 R^{ab} \, \varphi_a \, \varphi_b + 2\rho \, R - 12\varphi^a_{\ ab} \, \varphi^b +$$

$$- 3(\Box\varphi)^2 - 6 \, \varphi^{ab} \, \varphi_{ab}\,] + g^{\frac{1}{2}} \, u'' [-18 \, \rho \, \Box\varphi - 36\varphi^{ab} \, \varphi_a \, \varphi_b\,] - 12 \, g^{\frac{1}{2}} \, u''' \, \rho^2 \, , \qquad \text{Eq.1.24}$$

where u is an arbitrary scalar function of $\varphi$.

Note that $E^{ab}(L_{UC})$ is second-order in $g_{ab}$ and third order in $\varphi$, while $E(L_{UC})$ is third-order in $g_{ab}$ and fourth-order in $\varphi$. Of course, $E^a_{\ b}(L_{UC})$ and $E(L_{UC})$ are conformally invariant. I refer to $L_{UC}$ as the unexpected, conformally invariant

Lagrangian, although a better name might be, the ugly conformally invariant Lagrangian. $L_{UC}$ lacks the elegance of $L_{2C}$, $L_{3C}$ and $L_{4C}$ ; and its form can be simplified slightly by using the Schouten tensor, which, in a four-dimensional space, is given by

$$S^{ab} := \tfrac{1}{2}(R^{ab} - \tfrac{1}{6}\, g^{ab}\, R)\ .$$

The Schouten tensor has the property that

$$C^{abcd} = R^{abcd} + g^{ad}\, S^{bc} + g^{bc}\, S^{ad}\ - g^{ac}\, S^{bd}\ - g^{bd}\, S^{ac}\ .$$

$S^{ab}$ and $C^{abcd}$ can be used to simplify the form of Eq.1.23, but the improvement is insignificant.

An unfortunate aspect of $L_{UC}$ is that while it is second-order in $g_{ab}$ , it is third-order in $\varphi$. However, we can pull a divergence out of $L_{UC}$ to obtain

$$L_{UC} = L_{2UC} - (12\, g^{\frac{1}{2}}\, u\, \varphi^{k}\, \square\varphi\, )_{|k}\ , \qquad\qquad\text{Eq.1.25}$$

where

$$L_{2UC} := -\, 12\, g^{\frac{1}{2}}\, R^{hk}\, \varphi_{h}\, \varphi_{k}\ +\, 2\, g^{\frac{1}{2}}\, u\, R\, \rho - 6\, g^{\frac{1}{2}}\, u\, \varphi^{\,hk}\, \varphi_{hk} + 9\, g^{\frac{1}{2}}\, (\square\varphi)^{2}\ +$$

$$+\, 12\, g^{\frac{1}{2}}\, u'\, \rho\, \square\varphi\ . \qquad\qquad\text{Eq.1.26}$$

$L_{2UC}$ is second-order in $g_{ab}$ and $\varphi$, and I call it the second-order version of $L_{UC}$. Since $L_{2UC}$ is second-order, the formalism presented in Eqs.1.8-1.13 can be employed to determine its Euler-Lagrange tensor densities. In fact, that is how I arrived at Eqs.1.23 and 1.24. I shall have more to say about $L_{2UC}$ , and its relationship to the Lagrangians referred to as $L_{3}$ , $L_{4}$ and $L_{5}$ of Horndeski scalar theory, in the final section of this paper.

The primary purpose of this paper is to prove the following:

**Theorem:** In an orientable four-dimensional pseudo-Riemannian space, any conformally invariant, flat space compatible, scalar-tensor field theory, can have its field equations derived from the Lagrangian

$$L_C := L_{2C} + L_{3C} + L_{4C} + L_{UC} ,$$                     Eq.1.27

for a suitable choice of the functions k, p, b and u appearing in $L_{2C}$, $L_{3C}$, $L_{4C}$, and $L_{UC}$, respectively. These four conformally invariant, generating Lagrangians, are defined by Eqs. 1.5, 1.15, 1.18 and 1.22.■

The first thing you will notice about this theorem is that I demand that the spaces of interest must be orientable. This was done to guarantee that the Levi-Civita symbol, $\varepsilon^{abcd}$ , is a globally well-defined tensor density. However, since most of our work is confined to a coordinate domain, which is an orientable space, this assumption is not a severe restriction. Nevertheless, when we consider coordinate transformations, it will be assumed that the Jacobian is positive.

Another aspect of the theorem that you may have noticed, is that no mention is made of the differential order of the theories under consideration. That is because I shall prove that all theories which satisfy the assumptions of the theorem must have differential order less that or equal to four.

Since all (metric) tensor field theories, are trivially scalar-tensor field theories, we have the following immediate consequence of the theorem.

**Corollary:** In an orientable four-dimensional pseudo-Riemannian space, any conformally invariant, flat space compatible, tensor field theory, can have its field equations derived from the Lagrangian $L_{4C}$ given in Eq.1.18, with b being a constant. Thus the corresponding Euler-Lagrange tensor density must be a constant multiple of the Bach tensor density, which is presented in Eq.1.21.∎

The proof of the theorem breaks into two pieces. The first part involves proving the following

**Proposition:** In an orientable four-dimensional pseudo-Riemannian, any at most third-order, conformally invariant, flat-space compatible, scalar-tensor field theory, can have its field equations derived from the Lagrangian

$$L := L_{2C} + L_{3C} ,$$

for a suitable choice of the functions k and p appearing in $L_{2C}$ and $L_{3C}$, which are defined by Eqs.1.15 and 1.5, respectively.∎

The essential idea behind the proof of the Proposition, which is just a third-order version of the theorem, is to first compute the third-order part of the field equations. This can then be subtracted away from the initial field theory, leaving a second-order theory behind. Fortunately, all second-order scalar-tensor field theories have been constructed in [1], and the proof of the Proposition is easily completed.

The proof of the Theorem begins by showing that any theory that satisfies the assumptions, must be at most of fourth-order. Then the fourth-order parts of any such field theory will be computed, and subtracted away from the initial fourth-order theory, leaving us with a third-order one. The Proposition then takes care of the third-order part, and we are done proving the theorem. Now for the copious details, which I shall strive to make as comprehensible as possible.

## Section 2: Proof of the Proposition

The proof of the proposition will be accomplished by means of a lengthy series of lemmas. The first lemma will provide us with an easy way to spot conformally invariant scalar-tensor field theories. In fact, I already employed it to find the Lagrangian $L_{2C}$.

**Lemma 1:** Let $E^{ab}(L)$ and $E(L)$ be the Euler-Lagrange tensor densities of a scalar-tensor field theory. This theory will be conformally invariant if and only if $E^{ab}(L)$ is trace-free. If $E^{ab}(L)$ is trace-free, then L is conformally invariant up to a divergence.

**Proof:** ⇒ The Euler-Lagrange tensor densities of a scalar-tensor field theory are related by the identity (*see*, [1] or page 49 of [14])

$$E_a{}^b(L)_{|b} = \tfrac{1}{2}\, \varphi_a\, E(L) \,. \qquad\qquad Eq.2.1$$

If $g'_{ab} := e^{2\sigma} g_{ab}$ , we let $E_a{}^b(L)'$ and $E(L)'$ denote $E_a{}^b(L)$ and $E(L)$ built from $g'_{ab}$ and $\varphi$. Since Eq.2.1 is an identity, it is valid for every metric tensor and scalar field. Thus

$$E_a{}^b(L)'_{|'b} \ = \ \tfrac{1}{2} \, \varphi_a \, E(L)' \, , \qquad\qquad\qquad\qquad \text{Eq.2.2}$$

where "$_{|'}$" denotes covariant differentiation with respect to $g'_{ab}$.  Since

$$\Gamma'^r_{st} = \Gamma^r_{st} + (\sigma_s \, \delta_{tr} + \sigma_t \, \delta_s{}^r \ - \ g_{st} \, g^{rp} \, \sigma_p) \, , \qquad\qquad \text{Eq.2.3}$$

we easily find that

$$E_a{}^b(L)'_{|'b} \ = \ E_a{}^b(L)_{|b} \ - \ E_b{}^b(L) \, \sigma_a \, .$$

If we combine this equation with Eq.2.2, noting that $E(L)' = E(L)$, due to conformal invariance, we get

$$E_a{}^b(L)_{|b} - E_b{}^b(L) \, \sigma_a = \ \tfrac{1}{2} \, \varphi_a \, E(L) \, .$$

Hence $E_a{}^b(L)$ is trace-free due to Eq.2.1.

$\Leftarrow$The fact that $E_b{}^b(L) = 0$ implies, due to Proposition 2.1 in [11], that L is conformally invariant up to a divergence.  Thus we may write

$$L = L' + [\text{divergence}].$$

It is now apparent that $E_a{}^b(L) = E_a{}^b(L)'$ and $E(L) = E(L)'$. Thus $E_a{}^b(L)$ and $E(L)$ define a conformally invariant scalar-tensor field theory.$\blacksquare$

Now that we know how to recognize conformally invariant scalar-tensor field theories, let

$$L = L(g_{ab} \, ; g_{ab,c} \, ; . \, . \, .; \varphi; \varphi_{,c}; . \, . \, .) \qquad\qquad\qquad \text{Eq.2.4}$$

be a Lagrangian of arbitrary differential order in $g_{ab}$ and $\varphi$. Assume that

$$A^a{}_b := E^a{}_b(L) \quad \text{and} \quad B := E(L) \ ,\qquad\qquad \text{Eq.2.5}$$

are conformally invariant, and at most of third-order in the derivatives of $g_{ab}$ and $\varphi$. Thus due to Lemma 1 we know that L is conformally invariant up to a divergence and $A^{ab}$ is trace-free. The construction of $A^{ab}$ and B will proceed in two parts. First I shall construct the third-order parts in $A^{ab}$ and B. Let *L* denote the conformally invariant Lagrangian which yields these third-order terms. Then L − *L* will yield a second-order conformally invariant field theory. Since all second-order field theories were built in [1], we can pick out the conformally invariant second-order theories by looking at the traces of the Euler-Lagrange tensor densities presented in [1]. This will complete the proof of the Proposition.

We begin this construction by noting that due to Eq.2.1

$$A^{ab}{}_{|b} = \tfrac{1}{2}\,\varphi^a B \quad .\qquad\qquad\qquad \text{Eq.2.6}$$

Since $A^{ab}$ is at most third-order, $A^{ab}{}_{|b}$ will in general be of fourth-order. However, Eq.2.6 says that this is impossible. Consequently we have

$$\frac{\partial}{\partial g_{ij,klmn}}\,[A^{ab}{}_{|b}] = 0 \ \text{and} \ \frac{\partial}{\partial \varphi_{,klmn}}[A^{ab}{}_{|b}] = 0 \ .\qquad \text{Eq.2.7}$$

The fourth-order terms in $A^{ab}{}_{|b}$ are found in $A^{ab}{}_{,b}$, and thus Eq.2.7 gives us

$$\frac{\partial}{\partial g_{ij,klmn}}\left[\frac{\partial A^{ab}}{\partial g_{pq,rst}}\,g_{pq,rstb}\right] = 0 \ \ \text{and} \ \frac{\partial}{\partial \varphi_{,klmn}}\left[\frac{\partial A^{ab}}{\partial \varphi_{,pqr}}\,\varphi_{,pqrb}\right] = 0 \ .\qquad \text{Eq.2.8}$$

If we let

$$A^{ab;pq,rst} := \frac{\partial A^{ab}}{\partial g_{pq,rst}} \quad \text{and} \quad A^{ab;pqr} := \frac{\partial A^{ab}}{\partial \varphi_{,pqr}} \; ,$$

then we can rewrite Eq.2.8 as follows:

$$A^{a(k;|ij|;lmn)} = 0 \quad \text{and} \quad A^{a(k;lmn)} = 0 \; , \hspace{2cm} \text{Eq.2.9}$$

where parentheses around a string of digits denotes symmetrization over those digits, while the indices between vertical bars do not participate in the symmetrization process.

There are several other identities similar to those provided by Eq.2.9. which I shall now derive. These identities result from the coordinate and conformal invariance of $A^{ab}$ and B.

Assume that x and x' are two charts at an arbitrary point P of our space. Under the coordinate transformation $x'^i = x'^i(x^j)$, $A^{ab}$ transforms as follows:

$$A^{ab}(g'_{ij}; \ldots ; g'_{ij,klm} ; \varphi'; \ldots ; \varphi'_{,klm}) =$$

$$= (\det(J_s^r)) \, J'^a_c \, J'^b_d \, A^{cd}(g_{ij}; \ldots ; g_{ij,klm} ; \varphi; \ldots ; \varphi_{,klm} ) \hspace{1cm} \text{Eq.2.10}$$

where

$$\varphi' = \varphi, \; g'_{ij} = g_{rs} \, J_i^r \, J_j^s \; , \; J_i^r = \frac{\partial x^r}{\partial x'^i} \; \text{ and } \; J'^a_c = \frac{\partial x'^a}{\partial x^c} \quad . \hspace{1cm} \text{Eq.2.11}$$

If we replace $g'_{ij}$ and its derivatives with respect to $x'^j$ in Eq.2.10, using the expressions presented in Eq.2.11, we shall obtain an identity in $J_i^r$, $J_{ij}^{\;r}$, $J_{ijk}^{\;\;r}$ and $J_{ijkl}^{\;\;\;r}$, where, *e.g.,* $J_{ij}{}^r := \dfrac{\partial}{\partial x^{tj}} J_i{}^r$ . If we evaluate this identity at the point P , $J_i{}^r$ , $J_{ij}{}^r$ , $J_{ijk}{}^r$ and $J_{ijkl}{}^r$ are just an arbitrary collection of numbers, with $\det(J_i{}^r) > 0$, the obvious symmetries, such as $J_{ij}{}^r$ $=J_{(ij)}{}^r$ . So we can differentiate this identity with respect to $J_i{}^r$ , $J_{ij}{}^r$ , $J_{ijk}{}^r$ and $J_{ijkl}{}^r$ , and then evaluate these derivatives for the identity coordinate transformation, where $J_i{}^r =$ $\delta_i{}^r$ , and all of the other $J_{...}$'s are zero.  In this way we obtain four sets of identities called the (coordinate) invariance identities.  We shall only require the identities obtained by differentiating with respect to $J_{ijkl}{}^r$ .  The resulting identities for $A^{ab}$ and B  can be written as

$$A^{ab;i(j,klm)} = 0 \quad \text{and} \quad B^{:i(j,klm)} = 0 \ . \qquad\qquad \text{Eq.2.12}$$

Since $A^a{}_b$ and B are conformally invariant they must satisfy conformal invariance identities (*see,* du Plessis [21], for a general discussion of these identities). Under the conformal transformation $g_{ab} \to g'_{ab} := e^{2\sigma} g_{ab}$ , we find that

$$A^{ab}(\, e^{2\sigma} g_{ij}\,,\,(e^{2\sigma} g_{ij})_{,k}\,;\,(e^{2\sigma} g_{ij})_{,kl}\,;\,(e^{2\sigma} g_{ij})_{,klm}\,;\,\varphi;\,\varphi_{,k}\,;\,.\,.\,.\,;\,\varphi_{,klm}) =$$
$$= e^{-2\sigma}\, A^{ab}(\, g_{ij}\,;\,.\,.\,.\,;\,g_{ij,klm}\,;\,\varphi;\,.\,.\,.\,;\,\varphi_{,klm})\,, \qquad\qquad \text{Eq.2.13}$$

with a similar equation for B.  Upon differentiating Eq.2.13, and the counterpart for B, with respect to $\sigma_{,rst}$ , and then evaluating the result for $\sigma = 0$, we obtain

$$A^{ab;ij,klm}\, g_{ij} = 0 \quad \text{and} \ \ B^{:ij,klm}\, g_{ij} = 0 \ . \qquad\qquad \text{Eq.2.14}$$

We also have the identities

$$A^{ab;ij,klm}\, g_{ab} \ = \ 0 \quad \text{and} \ \ A^{ab;cde}\, g_{ab} = 0 \qquad\qquad \text{Eq.2.15}$$

which follow trivially from $A^{ab} g_{ab} = 0$ , but nevertheless, will be very useful.

Note that I have derived (coordinate) invariance identities, and conformal invariance identities, that exclusively involve the third-order derivatives of $A^{ab}$ and B. I did this because what we are trying to do now is determine the third-order parts of $A^{ab}$ and B, so that is why we need to find restrictions on these third-order terms.

To recapitulate the above work we have

**Lemma 2:** If $A^{ab}$ and B are the field tensor densities of a conformally invariant, third-order scalar-tensor field theory, then

$$A^{a(k;|ij|,lmn)} = 0 \; ; \; A^{a(i;jkl)} = 0 \; , \qquad\qquad \text{Eq.2.9}$$

$$A^{ab;i(j,klm)} = 0 \; ; \; B^{;i(j,klm)} = 0 \; , \qquad\qquad \text{Eq.2.12}$$

$$A^{ab;ij,klm} g_{ij} = 0 \; ; \; B^{;ij,klm} g_{ij} = 0, \qquad\qquad \text{Eq.2.14}$$

$$A^{ab;ij,klm} g_{ab} = 0 \; , \text{ and } \; A^{ab;cde} g_{ab} = 0 \; . \blacksquare \qquad\qquad \text{Eq.2.15}$$

To proceed further in our analysis of $A^{ab}$ and B we require a very powerful identity which is a generalization of a result presented by Aldersley (*see,* page 70 in [22], or [23]), where he investigated conformally invariant concomitants of the metric tensor. I present the scalar-tensor version of his identity in

**Lemma 3 (Aldersley's Identity):** Let $A^{ab}$ and B be the third-order scalar tensor concomitants defined by Eq.2.5. If $A^a_{\ b}$ and B are conformally invariant, then for every real number $\lambda > 0$,

$$\lambda^4 A^{ab}( g_{ij} ; \ldots ; g_{ij,klm} ; \varphi ; \ldots ; \varphi_{,klm} ) =$$

$$= A^{ab}( g_{ij}; \lambda g_{ij,k}; \lambda^2 g_{ij,kl}; \lambda^3 g_{ij,klm}; \varphi; \lambda \varphi_{,k}; \lambda^2 \varphi_{,kl}; \lambda^3 \varphi_{,klm}) \qquad \text{Eq.2.16}$$

and

$$\lambda^4 B( g_{ij}; \ldots ; g_{ij,klm} ; \varphi; \ldots ; \varphi_{,klm} ) =$$

$$= B( g_{ij}; \lambda g_{ij,k}; \lambda^2 g_{ij,kl}; \lambda^3 g_{ij,klm}; \varphi; \lambda \varphi_{,k}; \lambda^2 \varphi_{,kl}; \lambda^3 \varphi_{,klm}) . \qquad \text{Eq.2.17}$$

**Proof:** Since this lemma is so crucial in the proof of the Proposition and Theorem, I shall prove it in great detail.

Let P be an arbitrary point of our four-dimensional space, and let x be a chart at P with domain U. We define a second chart x' at P with domain U by: $x^i = \lambda x'^i$, where $\lambda > 0$, is a real number. Since $A^{ab}$ is a tensor density we know that

$$(\det(J_s{}^r))J'_c{}^a J'_d{}^b A^{cd}( g_{ij}; \ldots ; g_{ij,klm}; \varphi; \ldots ; \varphi_{,klm} ) =$$

$$= A^{ab}( g'_{ij}; \ldots ; g'_{ij,klm}; \varphi'; \ldots ; \varphi'_{,klm}) . \qquad \text{Eq.2.18}$$

Due to the tensor transformation laws we must have

$$g'_{ij} = \lambda^2 g_{ij} ; g'_{ij,k} = \lambda^3 g_{ij,k} ; \ldots ; g'_{ij,klm} = \lambda^5 g_{ij,klm}; \varphi' = \varphi; \varphi'_{,k} = \lambda \varphi_{,k} ; \ldots ; \varphi'_{,klm} = \lambda^3 \varphi_{,klm},$$

where the derivatives of the primed quantities are taken with respect to the chart x', and the derivatives of the unprimed quantities are taken with respect to the chart x. Using these transformation equations in Eq.2.18 we see that for every $\lambda > 0$ ,

$$\lambda^2 A^{ab}( g_{ij}; \ldots ; g_{ij,klm} ; \varphi ; \ldots ; \varphi_{,klm}) =$$

$$= A^{ab}( \lambda^2 g_{ij}; \lambda^3 g_{ij,k}; \lambda^4 g_{ij,kl}; \lambda^5 g_{ij,klm}; \varphi; \lambda \varphi_{,k}; \lambda^2 \varphi_{,kl}; \lambda^3 \varphi_{,klm}). \qquad \text{Eq.2.19}$$

I shall now show how the conformal invariance of $A^{ab}$ can be used to rewrite the right-hand side of Eq.2.19. To that end let $h_{ab}$ and $\psi$ be the x-components of a metric tensor and scalar field defined on a neighborhood of the point P. Under the conformal transformation $h_{ab} \rightarrow h'_{ab} := \lambda^2 h_{ab}$ we find that

$$A^{ab}(\lambda^2 h_{ij}; \ldots ; \lambda^2 h_{ij,klm}; \psi; \ldots ; \psi_{,klm}) = \lambda^{-2} A^{ab}(h_{ij}; \ldots ; h_{ij,klm}; \psi; \ldots ; \psi_{,klm}). \quad \text{Eq.2.20}$$

We now set

$$h_{ij} := g_{ij}(P) + \lambda g_{ij,k}(P)(x^k - x^k(P)) + \tfrac{1}{2}\lambda^2 g_{ij,kl}(P)(x^k - x^k(P))(x^l - x^l(P)) +$$

$$+ \tfrac{1}{6}\lambda^3 g_{ij,klm}(P)(x^k - x^k(P))(x^l - x^l(P))(x^m - x^m(P)),$$

and

$$\psi := \varphi(P) + \lambda \varphi_{,k}(P)(x^k - x^k(P)) + \tfrac{1}{2}\lambda^2 \varphi_{,kl}(P)(x^k - x^k(P))(x^l - x^l(P)) +$$

$$+ \tfrac{1}{6}\lambda^3 \varphi_{,klm}(P)(x^k - x^k(P))(x^l - x^l(P))(x^m - x^m(P)).$$

Since $h_{ij}(P) = g_{ij}(P)$, $h_{ij}$ is a well-defined metric tensor on a neighborhood of P. Using the above expressions for $h_{ij}$ and $\psi$ in Eq.2.20 we find that at P

$$A^{ab}(\lambda^2 g_{ij}; \lambda^3 g_{ij,k}; \lambda^4 g_{ij,kl}; \lambda^5 g_{ij,klm}; \varphi; \lambda\varphi_{,k}; \lambda^2\varphi_{,kl}; \lambda^3\varphi_{,klm}) =$$

$$= \lambda^{-2} A^{ab}(g_{ij}; \lambda g_{ij,k}; \lambda^2 g_{ij,kl}; \lambda^3 g_{ij,klm}; \varphi; \lambda\varphi_{,k}; \lambda^2\varphi_{,kl}; \lambda^3\varphi_{,klm}).$$

Combining this equation with Eq.2.19 demonstrates that Eq.2.16 is valid at P. Since P was an arbitrary point in our space, Eq.2.16 is valid in general.

The proof of the accuracy of Eq.2.17 is virtually identical to our proof of Eq.2.16, and will be omitted.∎

One should note that had we been working in an n-dimensional space, the $\lambda^4$

term on the left-hand sides of Eqs.2.16 and 2.17 would be replaced by $\lambda^n$.

Aldersley's identity is a very powerful functional equation which combines coordinate and conformal invariance. This identity will enable us to determine the basic functional form of $A^{ab}$ and $B$. But before we can use it to do that we require the following technical lemma which is due to Thomas [24].

**Lemma 4 (Thomas's Replacement Theorem):** If $T$ is a third-order tensorial concomitant of $g_{ab}$ and $\varphi$ of contravariant rank r, covariant rank s and weight w, which locally has the form

$$T^{a\cdots}_{\ \ b\ldots} = T^{a\cdots}_{\ \ b\ldots}(g_{ij};\ g_{ij,k};\ g_{ij,kl};\ g_{ij,klm};\ \varphi;\ \varphi_{,k};\ \varphi_{,kl};\ \varphi_{,klm})$$

then its value is unchanged if we replace its arguments by tensors in the manner indicated below

$$T^{a\cdots}_{\ \ b\ldots} =$$

$$= T^{a\cdots}_{\ \ b\ldots}(g_{ij};\ 0;\ \tfrac{1}{3}(R_{iklj} + R_{ilkj});\ \tfrac{1}{6}(R_{iklj|m} + R_{ilkj|m} + R_{ilmj|k} + R_{imlj|k} + R_{imkj|l} + R_{ikmj|l});$$

$$\varphi;\ \varphi_k;\ \varphi_{kl};\ \varphi_{(klm)})\ .\qquad\qquad\text{Eq.2.21}$$

**Proof:** This lemma was established for the pure metric case by Thomas in [24] (*see*, pages 96-100). The basic argument for the scalar-tensor case goes as follows. Let x be a chart of our space (which need not be 4-dimensional) at an arbitrary point P. Let y be the normal coordinate system at P determined by x, so that the tangent vectors $\partial/\partial x^i$ = $\partial/\partial y^i$, at P; and let $\gamma_{ab}$ denote the y-components of the metric tensor. Since T is a tensorial concomitant we know that at P

$$T^{a\cdots}{}_{b\ldots}(g_{ij};\ g_{ij,k};\ g_{ij,kl};\ g_{ij,klm};\ \varphi;\ \varphi_{,k};\ \varphi_{,kl};\ \varphi_{,klm}) =$$

$$= T^{a\cdots}{}_{b\ldots}(\gamma_{ij};\ \gamma_{ij,k};\ \gamma_{ij,kl};\ \gamma_{ij,klm};\ \varphi;\ \varphi_{,k};\ \varphi_{,kl};\ \varphi_{,klm}) \qquad\qquad \text{Eq.2.22}$$

where the derivatives on the right-hand side of Eq.2.22 are taken with respect to the chart y. We all know that at the pole of a normal coordinate system $\gamma_{ij,k} = 0$. All of the other derivatives of $\gamma$ and $\varphi$ with respect to y in Eq.2.22 are tensors at P. I establish this fact in Appendix B, following Thomas's arguments. In Appendix B, I also employ Thomas's work to explain why the values of these tensors are such that at P, Eq.2.22 assumes the form of Eq.2.21. Since P was an arbitrary point, our proof is now complete.∎

I would like to mention that in [24] Thomas assumes that the metric tensor is analytic. This just makes things simpler for him at various points, but it is not needed for what we are doing here.

With Lemmas 3 and 4 in hand, I can now construct the basic form of $A^{ab}$ and B.

**Lemma 5:** If $A^{ab} = E^{ab}(L)$, and $B = E(L)$, satisfy the assumptions of the Proposition, then

$$A^{ab} = \Psi^{abcdefhi}(g_{rs};\ \varphi)\, R_{cefd|h}\, \varphi_i\ + \Phi^{abcdef}(g_{rs};\varphi)\, \varphi_{(cde)}\, \varphi_f\ +$$

$$+\ a^{ab}(g_{rs};\ g_{rs,t};\ g_{rs,tu};\ \varphi;\ \varphi_{,t};\ \varphi_{,tu}) \qquad\qquad \text{Eq.2.23}$$

and

$$B = \Psi^{abcdef}(g_{rs}; \varphi) \, R_{acdb|e} \, \varphi_f + \Phi^{abcd}(g_{rs}; \varphi) \; \varphi_{(abc)} \, \varphi_d +$$

$$+ \, b(g_{rs}; g_{rs,t}; g_{rs,tu}; \varphi; \varphi_{,t}; \varphi_{,tu}) \qquad\qquad \text{Eq.2.24}$$

with $\Psi$ and $\Phi$ being tensor density concomitants of $g_{rs}$ and $\varphi$, while $a^{ab}$ and $b$ are second-order, scalar-tensor concomitants. $\Psi^{abcdefhi}$ has the same symmetries in the indices a,b,c,d,e,f,h as does $A^{ab;cd,efh}$, while $\Phi^{abcdef}$ has the same symmetries in the indices a,b,c,d,e as does $A^{ab;cde}$. Similar remarks apply to the symmetries of $\Psi^{abcdef}$ and $\Phi^{abcd}$.

**Proof:** Aldersley's identity for $A^{ab}$, Eq.2.16, tells us that

$$\lambda^4 \, A^{ab}(g_{ij}; g_{ij,k}; g_{ij,kl}; g_{ij,klm}; \varphi; \varphi_{,k}; \varphi_{,kl}; \varphi_{,klm}) =$$

$$= A^{ab}( \, g_{ij}; \lambda g_{ij,k}; \lambda^2 \, g_{ij,kl}; \; \lambda^3 \, g_{ij,klm}; \varphi; \lambda\varphi_{,k}; \lambda^2\varphi_{,kl}; \lambda^3 \, \varphi_{,klm} \, ) \, , \qquad \text{Eq.2.25}$$

for every $\lambda > 0$. Upon differentiating this equation with respect to $g_{cd,efh}$ we get

$$\lambda^4 \, A^{ab;cd,efh} = \lambda^3 \, A^{ab;cd,efh}( \, g_{ij}; \lambda g_{ij,k}; \lambda^2 g_{ij,kl}; \lambda^3 g_{ij,klm}; \varphi; \lambda \, \varphi_{,k}; \lambda^2 \, \varphi_{,kl}; \lambda^3 \, \varphi_{,klm} \, ) \, . \quad \text{Eq.2.26}$$

If we now differentiate Eq.2.26 with respect to $g_{rs,tuv}$ we obtain

$$A^{ab;cd,efh;rs,tuv} = \lambda^2 \, A^{ab;cd,efh;rs,tuv}(g_{ij}; \lambda \, g_{ij,k}; \lambda^2 \, g_{ij,kl}; \lambda^3 \, g_{ij,klm}; \varphi; \lambda \, \varphi_{,k}; \lambda^2 \, \varphi_{,kl}; \lambda^3 \, \varphi_{,klm}) \, .$$

Upon taking the limit of this equation as $\lambda \to 0^+$, recalling that our field theory is flat space compatible, we discover that

$$A^{ab;cd,efh;rs,tuv} = 0 \, . \qquad\qquad\qquad \text{Eq.2.27}$$

Hence $A^{ab}$ must be linear in third-order derivatives of $g_{ij}$ . In a similar way we can differentiate Eq.2.25 to demonstrate that

$$A^{ab;cd,efh;rst} = 0 \ , \ A^{ab;cd,efh;rs,tu} = 0 \ , \ A^{ab;cd,efh;rs} = 0, \ A^{ab;cd,efh;rs,t;uv,w} = 0, \qquad \text{Eq.2.28}$$

$$A^{ab;cd,efh;rs,t;u} = 0, \ \ A^{ab;cd,efh;r;s} = 0 \qquad \text{Eq.2.29}$$

$$A^{ab;cde;rst} = 0, \ A^{ab;cde;rs,tu} = 0, \ A^{ab;cde;rs} = 0, \ A^{ab;cde;rs,t;uv,w} = 0, \qquad \text{Eq.2.30}$$

$$A^{ab;cde;rs,t;u} = 0 \qquad A^{ab;cde;r;s} = 0 \ , \qquad \text{Eq.2.31}$$

where

$$A^{ab;rs,tu} := \frac{\partial A^{ab}}{\partial g_{rs,tu}}; \ \ A^{ab;rs} := \frac{\partial A^{ab}}{\partial \varphi_{,rs}} \ ; \ A^{ab;rs,t} := \frac{\partial A^{ab}}{\partial g_{rs,t}} \ \text{and} \ \ A^{ab;r} := \frac{\partial A^{ab}}{\partial \varphi_{,r}} \ .$$

Equations Eqs.2.27-2.31 allow us to deduce that

$$A^{ab} = \ \Psi^{abcdefhi} \ g_{cd,efh} \ \varphi_i + \Psi^{abcdefhijk} \ g_{cd,efh} \ g_{ij,k} + \Phi^{abcdef} \ \varphi_{,cde} \ \varphi_{,f} + \Phi^{abcdefhi}\varphi_{,cde} \ g_{fh,i} +$$

$$+ \ a^{ab}( \ g_{pq}; \ g_{pq,r}; \ g_{pq,rs} \ ; \ \varphi; \ \varphi_{,r}; \ \varphi_{,rs}) \ , \qquad \text{Eq.2.32}$$

where the $\Psi$ and $\Phi$ terms are concomitants of only $g_{pq}$ and $\varphi$. The coefficients $\Psi$ and $\Phi$ have numerous symmetries; *e.g.,*

$$\Psi^{abcdefhi} = \Psi^{(ab)cdefhi} = \ \Psi^{ab(cd)efhi} = \Psi^{abcd(efh)i} \qquad \text{Eq.2.33}$$

and, due to Lemma 2

$$\Psi^{a(b|cd|efh)i} = 0 \ , \ \ \Psi^{abc(defh)i} = 0 \ , \ g_{ab} \ \Psi^{abcdefhi} = 0, \ \text{and} \ \ g_{cd} \ \Psi^{abcdefhi} = 0 \ . \qquad \text{Eq.2.34}$$

Correspondingly,

$$\Phi^{abcdef} = \Phi^{(ab)cdef} = \Phi^{ab(cde)f} \ , \Phi^{a(bcde)f} = 0, \ \ \text{and} \ \ g_{ab} \ \Phi^{abcdef} = 0. \qquad \text{Eq.2.35}$$

It is easily seen that if you differentiate a scalar-tensor tensorial concomitant with respect to the highest order arguments of $g_{ab}$ and $\varphi$, appearing in it, you obtain tensorial concomitants. (The same is true if you differentiate with respect to φ.) Therefore $A^{ab;cd,efh}$ and $A^{ab;cde}$, are tensorial concomitants, and in view of Eq.2.32 they are only concomitants of $g_{ab}$, $g_{ab,c}$, $\varphi$ and $\varphi_{,c}$. Consequently

$A^{ab;cd,efh;i} = \Psi^{abcdefhi}$ ; $A^{ab;cd,efh;ij,k} = \Psi^{abcdefhijk}$ ; $A^{ab;cde;f} = \Phi^{abcdef}$ and $A^{ab;cde;fh,i} = \Phi^{abcdefhi}$

are tensorial concomitants of $g_{ab}$ and φ.

If we now apply the replacement theorem, Lemma 4, to Eq.2.32, we obtain Eq.2.23. The proof of Eq.2.24 is similar. Of course, we could apply lemma 4 to $a^{ab}$ and b in Eqs.2.23 and 2.24, but that will not be necessary at present.∎

Our next task is to compute the coefficient tensor densities in the expressions for $A^{ab}$ and B, given in Eqs.2.23 and 2.24. This is the task of Appendix C where I show how to prove

**Lemma 6:** In an orientable, 4-dimensional space, the general form of the tensor densities Ψ and Φ, appearing in Lemma 5 are given by

$$\Psi^{abcdefhi} =$$

$-2\alpha(\varphi)\,[(adhi)(bc)(ef) + (bdhi)(ac)(ef) + (achi)(bd)(ef) + (bchi)(ad)(ef) +$

$+\ (adei)(bc)(fh) + (bdei)(ac)(fh) + (acei)(bd)(fh) + (bcei)(ad)(fh) +$

$+\ (adfi)(bc)(eh) + (bdfi)(ac)(eh) + (acfi)(bd)(eh) + (bcfi)(ad)(eh)] +$

$+\ \alpha(\varphi)\,[(adhi)(be)(cf) + (bdhi)(ae)(cf) + (achi)(be)(df) + (bchi)(ae)(df) +$

$+\ (adei)(bh)(cf) + (bdei)(ah)(cf) + (acei)(bh)(df) + (bcei)(ah)(df) +$

$+ (adhi)(bf)(ce) + (bdhi)(af)(ce) + (achi)(bf)(de) + (bchi)(af)(de) +$

$+ (adei)(bf)(ch) + (bdei)(af)(ch) + (acei)(bf)(dh) + (bcei)(af)(dh) +$

$+ (adfi)(bh)(ce) + (bdfi)(ah)(ce) + (acfi)(bh)(de) + (bcfi)(ah)(de) +$

$+ (adfi)(be)(ch) + (bdfi)(ae)(ch) + (acfi)(be)(dh) + (bcfi)(ae)(dh)]$,　　Eq.2.36

$$\Phi^{abcdef} = 0, \quad \Psi^{abcdef} = 0,$$　　　　　　Eq.2.37

and

$$\Phi^{abcd} = g^{\frac{1}{2}} \beta(\varphi)[(ab)(cd) + (ac)(bd) + (ad)(bc)]$$　　Eq.2.38

where $\alpha = \alpha(\varphi)$ and $\beta = \beta(\varphi)$ are differentiable scalar fields, and, for typographical convenience, $(abcd) := \varepsilon^{abcd}$ and $(ef) := g^{ef}$. ∎

I shall now use Lemma 6 to build $A^{ab}$. From Eq.2.23 we see that we require expressions for $\Psi^{abcdefhi} R_{cefd|h} \varphi_i$ and $\Phi^{abcdef} \varphi_{(cde)} \varphi_f$. Using Eq.2.36 and Eq.2.37 we easily find that

$$\Psi^{abcdefhi} R_{cefd|h} \varphi_i = 24 \ \alpha[\varepsilon^{adhi}\varphi_i R^b_{d|h} + \varepsilon^{bdhi}\varphi_i R^a_{d|h}] \ ,$$　　　　Eq.2.39

and

$$\Phi^{abcdef} \varphi_{(cde)} \varphi_f \ = \ 0 \ .$$　　　　　　Eq.2.40

If we choose the function p appearing in $L_{3C} = p \ P$, so that $p' = 6\alpha$, we can use Eqs.1.6, 2.23 and 2.39 to deduce that

$$A^{ab} - E^{ab}(pP) = a^{ab}(g_{rs}; \ g_{rs,t}; \ g_{rs,tu}; \ \varphi; \ \varphi_{,t}; \ \varphi_{,tu}) \ .$$　　Eq.2.41

Since $A^a_b$ is a conformally invariant Euler-Lagrange tensor, we can conclude from Eq.2.41 that $a^a_b$ must be a conformally invariant, second-order Euler-Lagrange tensor density. At this juncture I would like to appeal to my work on second-order scalar-tensor field theories to complete the proof of the proposition. However, that is not quite possible yet. To do that we need to demonstrate that $B - E(pP)$ is second-order, and that has yet to be demonstrated. In fact due lemmas 4, 5 and 6 we can write

$B - E(pP) = g^{\frac{1}{2}} \beta (g^{ab} g^{cd} + g^{ac} g^{bd} + g^{ad} g_{bc}) \varphi_{abc} \varphi_d + \Phi_1^{abcd} \varphi_{ab} \varphi_{cd} + \Phi_2^{abcd} \varphi_{ab} \varphi_c \varphi_d +$

$+ \Phi_3^{abcd} \varphi_a \varphi_b \varphi_c \varphi_d + \Psi_1^{abcdef} R_{cabd} \varphi_{ef} + \Psi_2^{abcdef} R_{cabd} \varphi_e \varphi_f + \Psi^{abcdefhi} R_{cabd} R_{hefi}$ ,   Eq.2.42

where $\Psi_1, \Psi_2, \Psi, \Phi_1$ and $\Phi_2$ are tensorial concomitants of $g_{ab}$ and $\varphi$. These coefficient concomitants have various symmetries, most of which are obvious. A list of a few of these symmetries that we shall need shortly are:

$\Phi_1^{abcd} = \Phi_1^{(ab)cd} = \Phi_1^{ab(cd)} = \Phi_1^{cdab}; \Phi_2^{abcd} = \Phi_2^{(ab)cd} = \Phi_2^{ab(cd)}; \Phi_3^{abcd} = \Phi_3^{(abcd)}$ .    Eq.2.43

The first term on the right-hand side of Eq.2.42 reduces to

$$g^{\frac{1}{2}} \beta (\varphi^a_{ab} \varphi^b + 2\varphi^a_{ba} \varphi^b) = 3g^{\frac{1}{2}} \beta \varphi^a_{ab} \varphi + 2g^{\frac{1}{2}} \beta \varphi^a \varphi^b R_{ab} \ . \qquad \text{Eq.2.44}$$

The second term on the right-hand side of Eq.2.44 can be absorbed into the $\Psi_2^{abcdef} R_{cabd} \varphi_e \varphi_f$, in Eq.2.42. In view of the work done in Appendix C, and the symmetries presented in Eq.2.43, it is easy to show that

$$\Phi_1^{abcd} \varphi_{ab} \varphi_{cd} = g^{\frac{1}{2}} (\alpha_1 g^{ab} g^{cd} + \alpha_2 (g^{ac} g^{bd} + g^{ad} g^{bc})) \varphi_{ab} \varphi_{cd} =$$

$$= g^{\frac{1}{2}} \alpha_1 (\square \varphi)^2 + 2g^{\frac{1}{2}} \alpha_2 \ \varphi^{ab} \ \varphi_{ab} \qquad \text{Eq.2.45}$$

$$\Phi_2^{abcd} \varphi_{ab} \varphi_c \varphi_d = g^{\frac{1}{2}} (\alpha_3 g^{ab} g^{cd} + \alpha_4 (g^{ac} g^{bd} + g^{ad} g^{bc})) \varphi_{ab} \varphi_c \varphi_d =$$

$$= g^{\frac{1}{2}} \alpha_3 \rho \ \square \varphi + 2g^{\frac{1}{2}} \ \alpha_4 \ \varphi^{ab} \varphi_a \varphi_b \ , \qquad \text{Eq.2.46}$$

$$\Phi_3^{abcd} \varphi_a \varphi_b \varphi_c \varphi_d = g^{\frac{1}{2}} k \ \rho^2 \ , \qquad \text{Eq.2.47}$$

$$\Psi_1{}^{abcdef}R_{cabd}\varphi_{ef} = g^{\frac{1}{2}}\alpha_5 R\square\varphi + g^{\frac{1}{2}}\alpha_6 R^{ab}\varphi_{ab} \, , \qquad\qquad \text{Eq.2.48}$$

and

$$\Psi_2{}^{abcdef}R_{cabd}\varphi_e\varphi_f = g^{\frac{1}{2}}\alpha_7 R\rho + g^{\frac{1}{2}}\alpha_8 R^{ab}\varphi_a\varphi_b \, , \qquad\qquad \text{Eq.2.49}$$

where, $\alpha_1,...,\alpha_8$ and k are differentiable scalar functions of $\varphi$. If we now set $B:=$ B−E(pP), we can employ Eqs.2.42 and 2.44-2.49 to deduce that

$$B = g^{\frac{1}{2}}\beta\,\varphi^a{}_{ab}\,\varphi^b + g^{\frac{1}{2}}\alpha_1(\square\varphi)^2 + g^{\frac{1}{2}}\alpha_2\,\varphi^{ab}\,\varphi_{ab} + g^{\frac{1}{2}}\alpha_3\rho\square\varphi + g^{\frac{1}{2}}\alpha_4\varphi^{ab}\varphi_a\varphi_b + g^{\frac{1}{2}}\alpha_5 R\square\varphi +$$

$$+ g^{\frac{1}{2}}\alpha_6\,R^{ab}\varphi_{ab} + g^{\frac{1}{2}}\alpha_7\,\rho R + g^{\frac{1}{2}}\alpha_8\,R^{ab}\varphi_a\varphi_b + g^{\frac{1}{2}}k\rho^2 + \Psi^{abcdefhi}\,R_{cabd}\,R_{hefi} \, , \qquad \text{Eq.2.50}$$

where I have absorbed numerical constants into the $\alpha$'s, and renamed $\alpha_8 + 2\beta$, to be $\alpha_8$.

Let me briefly recapitulate where we are now. If we let $L:=$ L− pP, $A^{ab} := E^{ab}(L)$ and $B := E(L)$, then $A^{ab}$ and $B$ determine a conformally invariant scalar-tensor field theory which would be of second-order if we can demonstrate that $\beta$=0, in Eq.2.50. This is not any easy task, and has taxed my computational skills for quite a while, before I could find an argument that accomplished the objective. To that end we begin with a lemma that provides us with a remarkable property about scalar-tensor field theories in general.

**Lemma 7:** In an n-dimensional space, if L is a Lagrangian which is a concomitant of $g_{ab}$ and $\varphi$, along with their derivatives of arbitrary order, $A^{ab} := E^{ab}(L)$ and $B := E(L)$, then $E^{ab}(B) = -\dfrac{\partial A^{ab}}{\partial\varphi}$ and $E(B) = -\dfrac{\partial B}{\partial\varphi}$.

**Proof:** By definition we know that

$$B = E(L) = -\frac{\partial L}{\partial \varphi} + [\text{a divergence}] \ .$$

If we let the Euler-Lagrange operator E act on the above equation, then since

$$E[\text{a divergence}] = 0,$$

we get

$$E(B) = -E\left[\frac{\partial L}{\partial \varphi}\right] \ . \hspace{3cm} \text{Eq.2.51}$$

However, the differential operators $E^{ab}$ and E commute with $\frac{\partial}{\partial \varphi}$ and $\frac{\partial}{\partial g_{ab}}$ , and with no other partial derivatives. Hence Eq.2.51 shows that $E(B) = -\frac{\partial B}{\partial \varphi}$ . The proof that $E^{ab}(B) = -\frac{\partial A^{ab}}{\partial \varphi}$ , is similar.∎

At this point I would like to say that if we take an indefinite integral of Eq.2.51 with respect to $\varphi$, noting that $\int d\varphi$ commutes with E, then we get $E(\int Bd\varphi) = -B$. But unfortunately, $\int d\varphi$, acting on the partial derivative of L with respect to $\varphi$, gives you L, plus an arbitrary scalar density, which is a scalar-tensor concomitant independent of explicit $\varphi$ dependence. However, in practice one can usually circumvent this obstacle, to construct a second Lagrangian that yields $A^{ab}$ and B as its Euler-Lagrange tensor densities.

Lemma 7 is reassuring, for without something like it, we could use one conformally invariant scalar-tensor theory, derived from a Lagrangian, $\lambda$, to generate a whole family of such theories. This could be accomplished by looking at $E(\lambda)$, $E(E(\lambda))$, . . ., all of which are conformally invariant scalar densities, and hence can be taken as Lagrangians for conformally invariant scalar-tensor field theories. However, the Lemma tells us that this family is actually generated by

$$E(\lambda), \quad \frac{\partial E(\lambda)}{\partial \varphi}, \quad \frac{\partial^2 E(\lambda)}{\partial \varphi^2}, \ldots,$$

with corresponding field tensor densities, which are just repeated partial derivatives of $A^{ab}$ and $B$ with respect to $\varphi$. Hence this family is not really generating anything new.

In passing one should note that Lemma 7 is fairly obvious for the Lagrangians $L_{3C}$ and $L_{4C}$, where $E(L_{3C})$ and $E(L_{4C})$ are given by Eqs.1.7 and 1.20. But it is not so apparent for $L_{2C}$ and $L_{UC}$, where $E(L_{2C})$ and $E(L_{UC})$ are given by Eqs.1.17 and 1.24.

We shall now use Lemma 7 to prove that $\beta = 0$, in Eq.2.50, and in the process you will see how I found the Lagrangian $L_{UC}$.

**Lemma 8:** If $A^{ab}$ and $B$ are the Euler-Lagrange tensor densities given by $A^{ab} :=$ $E^{ab}(L-pP)$, and $B := E(L-pP)$, then the general form of $B$ is given by Eq.2.50 with $\beta=0$. Hence $A^{ab}$ and $B$ define a flat space compatible, second-order, conformally invariant, scalar-tensor field theory.

**Proof:** The scalar density $B$ given in Eq.2.50 is supposed to be conformally invariant. So let us examine what effect the conformal transformation $g_{ab} \rightarrow g'_{ab} := e^{2\sigma}g_{ab}$ has on $B$. But let's not be to hasty here, since this conformal transformation can yield quite a mess if one is not careful. The first thing we note is that we can disregard the term involving $\Psi^{abcdefhi} R_{cabd} R_{hefi}$, from our conformal transformation considerations here. This is so because it is second-order and has no terms involving derivatives of φ. So it just conformally transforms amongst itself, so to speak. We need to only concentrate on the differentiated φ terms in Eq.2.50.

Among the φ terms in Eq.2.50 of interest, we note that in the conformal transformation $B'$ of $B$, the only place where we shall encounter terms involving $R\sigma_h \varphi^h$ and $R^{ab}\varphi_a \sigma_b$, will be in a combination of the form

$$g^{\frac{1}{2}} [n_1 \alpha_5 R \varphi_h \sigma^h + n_2 \alpha_6 R^{ab} \varphi_a \sigma_b + n_3 \alpha_6 R \varphi_h \sigma^h],$$

where $n_1$, $n_2$, and $n_3$, are some numbers. This term must vanish identically if $B$ is to be conformally invariant. So it must vanish when $R_{ab} \neq 0$, but $R = 0$. This tells us that $\alpha_6 = 0$, since $\alpha_6$ is only a function of φ. Once $\alpha_6 = 0$, we see that $\alpha_5$ must also vanish. This greatly simplifies the conformal transformation of $B$. Another thing you should note about $B$ is that

$$g^{\frac{1}{2}} \beta\varphi^a{}_{ab} \varphi^b = g^{\frac{1}{2}}\beta( \Box\varphi)_{,b} \varphi^b.$$

This observation simplifies the conformal transformation of the β term in $B$. So if we now take the conformal transformation of $B$, denoting it by $B'$, we find that $B' = B$ if and only if

$$0 =$$

$$2\beta(-\sigma_a \varphi^a \Box\varphi + \sigma_{ab}\varphi^a\varphi^b - 2(\sigma_a\varphi^a)^2 + \varphi_{ab} \varphi^a\sigma^b) + 4\alpha_1(\varphi_a\sigma^a \Box\varphi +(\varphi_a \sigma^a)^2) + \rho\varphi_a\sigma^a(2\alpha_3 - \alpha_4)+$$

$$+ 2\alpha_2 \left(-2\varphi_{ab}\varphi^a\sigma^b + \varphi_a\sigma^a \,\square\varphi + \rho\sigma_a\sigma^a + (\varphi_a\,\sigma^a)^2\right) - 6\alpha_7\rho(\square\sigma + \sigma_a\,\sigma^a) +$$

$$+ \ \alpha_8 \left(2(\varphi_a\,\sigma^a)^2 - 2\sigma_{ab}\varphi^a\varphi^b - 2\rho\sigma_a\sigma^a - \rho\square\sigma\right) \qquad\text{Eq.2.52}$$

and

$$e^{-4\sigma}\,\Psi^{abcdefhi}R'_{cabd}\,R'_{hefi} = \Psi^{abcdefhi}R_{cabd}\,R_{hefi}\,, \qquad\text{Eq.2.53}$$

where Eq.2.53 is completely devoid of derivatives of $\varphi$, and not really that pertinent to our current quest. (I used the results of Appendix C to deduce how $\Psi$ must transform under a conformal transformation, since it must be built from all possible products of $g^{ab}$ and $\varepsilon^{cdef}$, with $g^{1/2}$ where necessary.) Upon comparing like terms in Eq.2.52 we deduce that $B' = B$ if and only if Eq.2.53 holds and

$$\alpha_1 = \tfrac{1}{4}\beta\,, \ \ \alpha_2 = \tfrac{1}{2}\beta\,, \ \ \alpha_4 = 2\alpha_3\,, \ \ \alpha_7 = -\tfrac{1}{6}\,\beta \ \ \text{and} \ \ \alpha_8 = \beta\,.$$

Using these restrictions on the $\alpha$ coefficients in Eq.2.50 shows us that for $B$ to be conformally invariant it must be given by

$$B = L_{UC} + g^{1/2}\alpha_4 \left(2\varphi^{ab}\varphi_a\,\varphi_b + \rho\square\varphi\right) + g^{1/2}k\rho^2 + \Psi^{abcdefhi}R_{cabd}\,R_{hefi} \qquad\text{Eq.2.54}$$

where $L_{UC}$ is given by Eq.1.22, with $u := -\tfrac{1}{12}\,\beta$, and $\Psi$ has had its components chosen so that the last term on the right-hand side of Eq.2.54 is conformally invariant. Using Eq.1.17 we can rewrite Eq.2.54 as follows

$$B = L_{UC} + E(\tfrac{1}{4}g^{1/2}\alpha_4\,\rho^2) + g^{1/2}(k -\tfrac{3}{4}\,\alpha_4{}')\rho^2 + \Psi^{abcdefhi}R_{cabd}\,R_{hefi}\,, \qquad\text{Eq.2.55}$$

where now, "'", once again, denotes a derivative with respect to $\varphi$, and $\Psi$ has been built so that the last term on the right-hand side of Eq.2.55 is conformally invariant.

From Lemma 7 we know that $E(B) = -B'$. But due to Eq.1.24 we know that $E(L_{UC})$ is of fourth-order in $\varphi$, unless $u = -^1/_{12}\beta = 0$. This observation finishes the proof of the Lemma.∎

Due to Lemma 8 we have, at long last, reduced our original third-order scalar-tensor problem to a second-order one. Using Lemma 1 to analyze the second-order scalar tensor field theories presented in [1], we discover that the only flat space compatible, conformally invariant, second-order scalar-tensor field theory, is generated by the Lagrangian $L_{2C}$. This observation, combined with Lemma 8 shows that

$$A^{ab} := E^{ab}(L) = E^{ab}(L_{2C} + L_{3C}) \text{ and } B := E(L) = E(L_{2C} + L_{3C}),$$

for suitable choices of the functions k and p appearing in $L_{2C}$ and $L_{3C}$. This completes the proof of our Proposition.

In the next section I shall utilize, and generalize, the machinery developed so far, to prove the Theorem. Fortunately, due to what we have gone through, the proof will not be as long as the Proposition's proof.

**Section 3: The Proof of the Theorem**

I shall begin this section by proving that any conformally invariant, flat space compatible, scalar-tensor field theory, must have differential order less than or equal to four. To that end if $A^{ab}$ and B are of $k^{th}$ order in the field variables, then we shall express their functional form as follows:

$$A^{ab} = A^{ab}(\ g;\ \partial g;\ \ldots;\ \partial^k g;\ \varphi;\ \partial\varphi;\ldots;\ \partial^k\varphi),$$

$$B\ =\ \ B(\ g;\ \partial g;\ldots;\ \partial^k g;\ \varphi;\ \ \partial\varphi;\ldots;\ \ \partial^k\varphi)\ .$$

Using this notation I can now state

**Lemma 9 (Aldersley's General Scalar-Tensor Identity):** Let $A^{ab}$ and $B$ be the $k^{th}$ order, scalar-tensor Euler-Lagrange tensor densities, generated in an n-dimensional pseudo-Riemmanian space, by the Lagrangian L. If $A^a_{\ b}$ and $B$ are conformally invariant, then for every real number $\lambda > 0$

$$\lambda^n A^{ab}(g;\ \partial g;\ldots;\ \partial^k g;\ \varphi;\ \partial\varphi;\ldots;\ \partial^k\varphi) =$$

$$= A^{ab}(\ g;\ \lambda\partial g;\ \ldots;\ \lambda^k\ \partial^k g;\ \varphi;\ \lambda\partial\varphi;\ \ldots;\ \lambda^k\partial^k\varphi) \qquad\qquad \text{Eq.3.1}$$

and

$$\lambda^n B(g;\ \partial g;\ \ldots;\ \partial^k g;\ \varphi;\ \partial\varphi;\ldots;\ \partial^k\varphi) =$$

$$=\ \ B(\ g;\ \lambda\partial g;\ \ldots;\ \lambda^k\partial^k g;\ \varphi;\ \lambda\partial\varphi;\ \ldots;\ \lambda^k\ \partial^k\varphi)\ , \qquad\qquad \text{Eq.3.2}$$

where there is no sum over the repeated superscripts k.

**Proof:** The proof of this lemma is virtually identical to the proof of Lemma 3, and will be omitted.∎

As an immediate consequence of this lemma we have

**Lemma 10:** In an n-dimensional   pseudo-Riemannian space, any $k^{th}$ order, conformally invariant, flat space compatible, scalar-tensor field theory, must have $k \le n$. In particular, in a 4-dimensional pseudo-Riemannian space, all flat space compatible, conformally invariant, scalar-tensor field theories, must have order $\leq 4$.

**Proof:** Let $A^{ab}$ and B be the field tensor densities of the theory in question. Upon differentiating Eqs.3.1 and 3.2 with respect to $\partial^k g$ we get

$$\lambda^n \frac{\partial A^{ab}}{\partial(\partial^k g)} = \lambda^k \frac{\partial A^{ab}}{\partial(\partial^k g)} ( g; \lambda\partial g; \ldots; \lambda^k\partial^k g; \varphi; \lambda\partial\varphi; \ldots; \lambda^k\partial^k\varphi) \qquad \text{Eq.3.3}$$

and

$$\lambda^n \frac{\partial B}{\partial(\partial^k g)} = \lambda^k \frac{\partial B}{\partial(\partial^k g)} ( g; \lambda\partial g; \ldots; \lambda^k\partial^k g; \varphi; \lambda\partial\varphi; \ldots; \lambda^k\partial^k\varphi) . \qquad \text{Eq.3.4}$$

If $k > n$, we multiply Eqs. 3.3 and 3.4 by $\lambda^{-n}$. Then upon taking the limit as $\lambda \to 0^+$ in the resulting equations, we find that

$$\frac{\partial A^{ab}}{\partial(\partial^k g)} = 0 , \quad \frac{\partial B}{\partial(\partial^k g)} = 0 .$$

Hence $A^{ab}$ and B must be independent of $k^{th}$ order derivatives of $g_{ab}$. In a similar way we can prove that $A^{ab}$ and B must be independent of $k^{th}$ order derivatives of $\varphi$. Consequently, the differential order of the field theory generated by $A^{ab}$ and B must be $\leq n$. ∎

Throughout the remainder of this section we shall confine our attention to 4-dimensional spaces, and it will be assumed that $A^{ab}$ and B satisfy the assumptions of the Theorem. We would now like to use Aldersley's General Scalar-Tensor Identity to determine the functional form of $A^{ab}$ and B. To assist in that endeavor we need the analog of Lemma 2, which is

**Lemma 11:** If $A^{ab}$ and B are the field tensor densities of a fourth-order, conformally invariant, scalar-tensor field theory, then

$$A^{a(b;|cd|,efhi)} = 0, \quad A^{a(b;cdef)} = 0 , \qquad\qquad\qquad \text{Eq.3.5}$$

$$A^{ab;c(d,efhi)} = 0, \quad B^{;c(d,efhi)} = 0 , \qquad\qquad\qquad \text{Eq.3.6}$$

$$A^{ab;cd,efhi} \, g_{cd} = 0, \quad B^{;cd,efhi} \, g_{cd} = 0 , \qquad\qquad\qquad \text{Eq.3.7}$$

and

$$A^{ab;cd,efhi} \, g_{ab} = 0, \quad A^{ab;cdef} \, g_{ab} = 0 , \qquad\qquad\qquad \text{Eq.3.8}$$

where

$$A^{ab;cd,efhi} := \frac{\partial A^{ab}}{\partial g_{cd,efhi}} \; ; A^{ab;cdef} := \frac{\partial A^{ab}}{\partial \varphi_{,cdef}} \text{ and } \quad B^{;cd,efhi} := \frac{\partial B}{\partial g_{cd,efhi}} .$$

**Proof:** Since the proof of this lemma is similar to the proof of Lemma 2, I shall only quickly sketch the details. We know from Eq.2.1 that

$$A^{ab}{}_{|b} = \tfrac{1}{2} \, \varphi^a \, B .$$

Since $A^{ab}$ and B are fourth-order we must have

$$\frac{\partial A^{ab}{}_{|b}}{\partial g_{cd,efhij}} = 0 \quad \text{and} \quad \frac{\partial A^{ab}{}_{|b}}{\partial \varphi_{,cdefh}} = 0 \quad .$$

This gives us Eq.3.5.

Since $A^{ab}$ and B are tensor densities we can examine how they transform under a coordinate transformation. Doing this will give us an identity in the Jacobian matrix $J_b{}^a$ and its derivatives. Upon differentiating that identity with respect to the highest occurring derivative of $J_b{}^a$ in that equation we obtain Eq.3.6.

Eq.3.7 is obtained by examining the effect of the conformal transformation $g_{ab} \rightarrow g'_{ab} = e^{2\sigma}g_{ab}$ on $A^a_{\ b}$ and B. This gives rise to an identity in $\sigma$; $\sigma_{,a}$; . . . ; $\sigma_{,abcd}$. When these identities are differentiated with respect to $\sigma_{,abcd}$ , Eq.3.7 results.

Lastly, Eq.3.8 is a trivial consequence of the fact that Lemma 1 tells us that $A^{ab}g_{ab} = 0$, and hence

$$\frac{\partial(A^{ab}g_{ab})}{\partial g_{cd,efhi}} = 0 \quad \text{and} \quad \frac{\partial(A^{ab}g_{ab})}{\partial \varphi_{,cdefh}} = 0 \ . \blacksquare$$

Continuing to proceed just as we did in Section 2, we can construct the basic functional form of $A^{ab}$ and B using Lemmas 9 and 11. This result is presented in

**Lemma 12:** If $A^{ab} = E^{ab}(L)$, and $B = E(L)$, satisfy the assumptions of the Theorem then

$$A^{ab} = \Psi^{abcdefhi}g_{cd,efhi} + \Phi^{abcdef}\varphi_{,cdef} + a^{ab}(g_{rs}; \ldots; g_{rs,tuv}; \varphi; \ldots; \varphi_{,tuv}) \ , \qquad \text{Eq.3.9}$$

and

$$B = \Psi^{abcdef}g_{ab,cdef} + \Phi^{abcd}\varphi_{,abcd} + b(g_{rs}; \ldots; g_{rs,tuv}; \varphi; \ldots; \varphi_{,tuv}) \ , \qquad \text{Eq.3.10}$$

where $\Psi$ and $\Phi$ are tensor density concomitants of $g_{ab}$ and $\varphi$. The coefficient concomitants have the following symmetries:

$$\Psi^{abcdefhi} = \Psi^{(ab)cdefhi} = \Psi^{ab(cd)efhi} = \Psi^{abcd(efhi)};$$

$$\Psi^{a(b|cd|efhi)} = 0, \ \Psi^{abc(defhi)} = 0, \ g_{ab}\Psi^{abcdefhi} = 0 \ , \ g_{cd}\Psi^{abcdefhi} = 0;$$

$$\Phi^{abcdef} = \Phi^{(ab)cdef} = \Phi^{ab(cdef)}, \ \Phi^{a(bcdef)} = 0, \ g_{ab}\Phi^{abcdef} = 0;$$

$$\Psi^{abcdef} = \Psi^{(ab)cdef} = \Psi^{ab(cdef)}; \ g_{ab}\Psi^{abcdef} = 0 \ , \ \Psi^{a(bcdef)} = 0 \text{ and } \Phi^{abcd} = \Phi^{(abcd)} \ . \blacksquare$$

At this point we could use the fourth-order version of Thomas's Replacement Theorem to "firm up" the functional form of $A^{ab}$ and B in Eqs.3.9 and 3.10. But that won't be necessary yet. What we really need is the next lemma which provides us with the functional form of the $\Psi$'s and $\Phi$'s.

**Lemma 13:** In a four-dimensional space, the coefficient tensor densities $\Phi^{abcd}$, $\Phi^{abcdef}$, $\Psi^{abcdef}$ and $\Psi^{abcdefhi}$, are given by

$$\Phi^{abcd} = g^{\frac{1}{2}}\beta(g^{ab}\,g^{cd} + g^{ac}\,g^{bd} + g^{ad}\,g^{bc})\,,\; \Phi^{abcdef} = 0\,,\; \Psi^{abcdef} = 0\,, \text{ and}$$

$$\Psi^{abcdefhi} =$$

$g^{\frac{1}{2}}\alpha\{(ab)(ce)(di)(fh) + (ab)(ci)(de)(fh) + (ab)(cf)(di)(eh) + (ab)(ci)(df)(eh) +$

$+ (ab)(ch)(di)(ef) + (ab)(ci)(dh)(ef) + (ab)(ce)(dh)(fi) + (ab)(ch)(de)(fi) +$

$+ (ab)(cf)(dh)(ei) + (ab)(ch)(df)(ei) + (ab)(ce)(df)(hi) + (ab)(cf)(de)(hi) +$

$+ (ae)(bi)(cd)(fh) + (ai)(be)(cd)(fh) + (af)(bi)(cd)(eh) + (ai)(bf)(cd)(eh) +$

$+ (ah)(bi)(cd)(ef) + (ai)(bh)(cd)(ef) + (ae)(bh)(cd)(fi) + (ah)(be)(cd)(fi) +$

$+ (af)(bh)(cd)(ei) + (ah)(bf)(cd)(ei) + (ae)(bf)(cd)(hi) + (af)(be)(cd)(hi) +$

$+ (ae)(bf)(ch)(di) + (af)(be)(ch)(di) + (ae)(bf)(ci)(dh) + (af)(be)(ci)(dh) +$

$+ (ae)(bh)(cf)(di) + (ah)(be)(cf)(di) + (ae)(bh)(ci)(df) + (ah)(be)(ci)(df) +$

$+ (af)(bh)(ce)(di) + (ah)(bf)(ce)(di) + (af)(bh)(ci)(de) + (ah)(bf)(ci)(de) +$

$+ (ae)(bi)(cf)(dh) + (ai)(be)(cf)(dh) + (ae)(bi)(ch)(df) + (ai)(be)(ch)(df) +$

$+ (af)(bi)(ch)(de) + (ai)(bf)(ch)(de) + (ai)(bf)(ce)(dh) + (af)(bi)(ce)(dh) +$

$+ (ah)(bi)(ce)(df) + (ai)(bh)(ce)(df) + (ah)(bi)(cf)(de) + (ai)(bh)(cf)(de) +$

$-\,^{3}/_{2}[(ac)(bi)(de)(fh) + (ai)(bc)(de)(fh) + (ad)(bi)(ce)(fh) + (ai)(bd)(ce)(fh) +$

+ (ac)(bi)(df)(eh) + (ai)(bc)(df)(eh) + (ad)(bi)(cf)(eh) + (ai)(bd)(cf)(eh) +

+ (ac)(bi)(dh)(ef) + (ai)(bc)(dh)(ef) + (ad)(bi)(ef)(ch) + (ai)(bd)(ch)(ef) +

+ (ac)(bh)(de)(fi) + (ah)(bc)(de)(fi) + (ad)(bh)(ce)(fi) + (ah)(bd)(ce)(fi) +

+ (ac)(bh)(df)(ei) + (ah)(bc)(df)(ei) + (ad)(bh)(cf)(ei) + (ah)(bd)(cf)(ei) +

+ (ac)(bf)(de)(hi) + (af)(bc)(de)(hi) + (ad)(bf)(ce)(hi) + (af)(bd)(ce)(hi) +

+ (ac)(be)(df)(hi) + (ae)(bc)(df)(hi) + (ad)(be)(cf)(hi) + (ae)(bd)(cf)(hi) +

+ (ac)(bf)(dh)(ei) + (af)(bc)(dh)(ei) + (ad)(bf)(ch)(ei) + (af)(bd)(ch)(ei) +

+ (ac)(be)(dh)(fi) + (ae)(bc)(dh)(fi) + (ad)(be)(ch)(fi) + (ae)(bd)(ch)(fi) +

+ (ac)(be)(di)(fh) + (ae)(bc)(di)(fh) + (ad)(be)(ci)(fh) + (ae)(bd)(ci)(fh) +

+ (ac)(bf)(di)(eh) + (af)(bc)(di)(eh) + (ad)(bf)(ci)(eh) + (af)(bd)(ci)(eh) +

+ (ac)(bh)(di)(ef) + (ah)(bc)(di)(ef) + (ad)(bh)(ci)(ef) + (ah)(bd)(ci)(ef)] +

+ 6[(ac)(bd)(ef)(hi) + (ad)(bc)(ef)(hi) + (ac)(bd)(eh)(fi) + (ad)(bc)(eh)(fi) +

+ (ac)(bd)(ei)(fh) + (ad)(bc)(ei)(fh)] +

− 4[(ab)(cd)(ef)(hi) + (ab)(cd)(eh)(fi) + (ab)(cd)(ei)(fh)]}

where α and β are scalar functions of φ.

**Proof:** Simply use Weyl's result on concomitants of $g_{ab}$ and φ, as described in Appendix C, to grind out these formulas. It is a time consuming, but straight-forward task.∎

    At this point the end of the Theorem's proof is now in sight. Using Lemmas 12

and 13 we see that $A^{ab}$ is fourth-order in $g_{ab}$ and third-order in $\varphi$. Moreover, the fourth-order metric part has only one arbitrary function of $\varphi$ in it. Now $E^{ab}(L_{4C})$ has only one arbitrary function of $\varphi$, $b(\varphi)$, and it is multiplying the fourth-order terms in $E^{ab}(L_{4C})$. Evidently b can be chosen so that the fourth-order terms in $A^{ab}$ are equal to those in $E^{ab}(L_{4C})$. Thus if we now consider the Lagrangain $L := L - L_{4C}$, we see that our original fourth-order problem has been reduced to one in which $E^{ab}(L)$, is third-order, and $E(L)$ is at most fourth-order. But Lemmas 12 and 13 tell us that $E(L)$ is devoid of fourth-order $g_{ab}$, and the fourth-order $\varphi$ terms in $E(L)$ are given by $\Phi^{abcd}\varphi_{,abcd}$. The replacement theorem has this term replaced by $\Phi^{abcd}\varphi_{(abcd)}$. Hence we may write

$$E(L) = g^{\frac{1}{2}}\beta(g^{ab}\,g^{cd} + g^{ac}\,g^{bd} + g^{ad}\,g^{bc})\varphi_{abcd} + \text{third order terms}$$

and so

$$E(L) = 3g^{\frac{1}{2}}\beta\square\square\varphi + \text{third order terms},$$

where I have made use of the fact that $\Phi^{abcd}$ is totally symmetric. Thus we see from Eq.1.24, that the fourth-order $\varphi$ terms in $E(L)$ are the same as those in $E(L_{UC})$, when we take $u = \frac{1}{2}\beta$. Consequently all of the fourth-order terms in $E^{ab}(L)$ and $E(L)$, are in $E^{ab}(L_{4C} + L_{UC})$ and $E(L_{4C} + L_{UC})$, for suitable choices of b and u. As a result, if we now consider the Lagrangian $L - L_{4C} - L_{UC}$, we see that it yields a third-order, flat space compatible, conformally invariant, scalar-tensor field theory. Hence our Proposition tells us that the theory generated by $L - L_{4C} - L_{UC,}$ can also be generated by $L_{2C} + L_{3C,}$ for a suitable choice of the functions k and p.  At long last, this observation completes the proof of the Theorem.

**Section 4: Concluding Remarks**

Now that we have this wonderful Theorem that presents us with all conformally invariant, flat space compatible, scalar-tensor field theories, in a four-dimensional space, what do we do next?  Well, the first thing we could  do is check to see which of the Lagrangians $L_{2C}$, $L_{3C}$, $L_{4C}$ and $L_{UC}$  yield field theories  which avoid an Ostrogradsky type instability, as described in [25].  Evidently, $L_{2C}$, will not cause any problems, and $L_{4C}$ has got to cause trouble, by virtue of the fact that $E^{ab}(L_{4C})$ is fourth-order in $g_{ab}$.  I leave it to those who are more knowledgeable about instabilities to determine whether  $L_{3C}$ yields a stable theory.  Before I discuss the instability problem for $L_{UC}$, I would like to  make a few more remarks about it.

In [11] I  commented about how fond I was of the Lagrangian $L_{3C}$.  Well, now $L_{UC}$ is the "apple of my eye," for several reasons.  First of all, unlike $L_{2C}$, $L_{3C}$ and $L_{4C}$; $L_{UC}$ has derivatives of both $g_{ab}$ and $\varphi$ in it.  Secondly, $E^{ab}(L_{UC})$ is second-order in the derivatives of $g_{ab}$, while $E^{ab}(L_{3C})$ and $E^{ab}(L_{4C})$, are third and fourth-order in derivatives of $g_{ab}$, respectively, with $E^{ab}(L_{2C})$ being devoid of derivatives of $g_{ab}$.  Ostensibly, one draw back of $L_{UC}$ is that $E(L_{UC})$ is third-order in $g_{ab}$ and fourth-order in $\varphi$, but I do not regard this to be a problem. For recall that due to Eq.2.1, we know that

$$E^{ab}(L_{UC})_{|b} = \tfrac{1}{2}\varphi^a E(L_{UC}).$$

Thus when working in a vacuum the $E(L_{UC})$ equation is superfluos, so who cares about its differential order. For all practical purposes, $L_{UC}$ defines a conformally invariant scalar-tensor field theory, that is second-order in $g_{ab}$, and third-order in $\varphi$. I shall now show how $L_{UC}$ is related to the Lagrangians of Horndeski Scalar Theory.

In terms of conventional notation, the Lagrangians $L_3$, $L_4$ and $L_5$ of Horndeski Scalar Theory are

$$L_3 := g^{\tfrac{1}{2}} G_3 \square\varphi \; ; \quad L_4 := g^{\tfrac{1}{2}} G_4 R - 2g^{\tfrac{1}{2}} G_{4,\rho}((\square\varphi)^2 - \varphi^{ab}\varphi_{ab}) \, , \text{ and}$$

$$L_5 := g^{\tfrac{1}{2}} G_5 G^{ab}\varphi_{ab} - \tfrac{1}{3} g^{\tfrac{1}{2}} G_{5,\rho}((\square\varphi)^3 - 3\varphi^{ab}\varphi_{ab}\square\varphi + 2\varphi^a{}_b\,\varphi^b{}_c\,\varphi^c{}_a) \, ,$$

where $G_3$, $G_4$ and $G_5$ are scalar functions of $\varphi$ and $\rho$, with ",$\rho$" denoting a partial derivative with respect to $\rho$. Let us now chose $G_3 = 12u'\rho$, $G_4 = -4u\rho$, and $G_5$ to be independent of $\rho$, and such that $G_5' = 12u$. If we let $\Lambda_3$, $\Lambda_4$ and $\Lambda_5$ denote what $L_3$, $L_4$ and $L_5$ become for the aforementioned choices of $G_3$, $G_4$ and $G_5$, then we find that

$$\Lambda_3 + \Lambda_4 + \Lambda_5 + \Lambda - (g^{\tfrac{1}{2}}[\textstyle\int 12ud\varphi]G^{ab}\varphi_a)_{|b} = L_{2UC}, \qquad\qquad \text{Eq.4.1}$$
where
$$\Lambda := g^{\tfrac{1}{2}}((\square\varphi)^2 + 2\,\varphi^{ab}\varphi_{ab}). \qquad\qquad\qquad\qquad\qquad \text{Eq.4.2}$$

Due to Eqs.4.1 and 4.2, we see that the field theories generated by $L_{UC}$, or equivalently by $L_{2UC}$, differs from a second-order scalar-tensor field theory owing to the Lagrangian $\Lambda$, which is quadratic in the second derivatives of $\varphi$. Thus all of the higher-order terms in $E^{ab}(L_{UC})$ and $E(L_{UC})$, owe their origin to $\Lambda$. Beyond Horndeski Theories, such as those generated by $\Lambda$, were studied extensively by Langlois and Noui [26]. Their analysis shows that the Lagrangian $\Lambda_4 + \Lambda$, generates a theory with a degenerate kinetic matrix, and hence it is likely to avoid Ostrogradsky type singularities. However, $L_{2UC}$ is more general than $\Lambda_4 + \Lambda$. For the case of $L_{2UC}$ they go on to say, "restricted combinations of Horndeski's Lagrangians with" Lagrangians like $\Lambda$, "lead to degenerate theories and thus are presumably free of Ostrogradsky instablities," (quote from page 17 of [26]). These remarks suggest that there is a good reason to hope that Lagrangian $L_{2UC}$ is stable. However, those hopes are dashed by the work of Achour, *et al.,*[27], who show (*see* page 16 of [27]), that combinations of quartic beyond Horndeski Lagrangians, such as $\Lambda$, with all other Horndeski Lagrangians other than those of type $L_4$, will lead to theories with instabilities. If this is so, then $L_{UC}$ will probably be viewed by some as an "unphysical" choice of Lagrangian.

At this time it should be noted that in $\Lambda_3$, u' appears. Thus if we choose $u = \mu$ (a constant), then $\Lambda_3$ would vanish. Let $L_{2\mu C}$ denote $L_{2UC}$ when $u = \mu$. I shall also set $G_5 = 12\mu\varphi$ in $L_{2\mu C}$. In terms of the nomenclature employed in[26], $L_{2\mu C}$ is the sum of a quartic and quintic Horndeski Lagrangian, with a quartic Beyond Horndeski Lagrangian. Due to [27] such a Lagrangian would in general have instabilities. But perhaps a closer look at $L_{2\mu C}$ is warranted, due to the highly simple form of $G_5$ in it. My hope is that $L_{2\mu C}$ will be free of Ostrogradsky instabilites. But even if it isn't, it is still a very interesting Lagrangian.

Let's stop thinking about instabilities in conformally invariant scalar-tensor field theories, and consider another problem. What possible good is a conformally invariant scalar-tensor field theory anyway? The gravitational field outside of the sun is clearly not conformally invariant. So it is evident that gravity, at all times throughout the Universe, can not be described by a conformally invariant scalar-tensor field theory. Now before we completely trash conformally invariant scalar-tensor field theories, let's consider the Einstein-Maxwell field equations. The Lagrangian for the electromagnetic field is

$$L_{EM} := \tfrac{1}{4}\gamma \ g^{\frac{1}{2}} \ F^{ab} \ F_{ab}$$

where $\gamma$ is a constant, and $F_{ab} := \psi_{a,b} - \psi_{b,a}$ with $\psi_a$ denoting the vector potential. $L_{EM}$ is conformally invariant, and hence $E^a_{\ b}(L_{EM})$ and $E^a(L_{EM})$ are also conformally invariant. The Einstein-Maxwell field equations are obtained from the Lagrangian $g^{\frac{1}{2}}R + L_{EM}$. Thus we see that the pure conformal invariance of the electromagnetic field is broken by the introduction of local gravity, which is manifested by $g^{\frac{1}{2}}R$.

This suggests that one might regard a conformally invariant scalar-tensor field theory as describing what gravity would be like in the Universe before local effects arose, or if local effects could be turned off. But when local effects arise, they break the conformal invariance of the theory. Consequently, the Lagrangian for gravity throughout the Universe, when local effects have manifested themselves, could have the form

$$L_G = g^{½}R + L_C ,$$ Eq.4.3

where $L_C$ is one of our conformally invariant, flat space compatible, scalar-tensor Lagrangians. Due to Lovelock's work [28], we know $g^{½}( R + \lambda)$, yields the most general second-order metric tensor field equations in a 4-dimensional space, where $\lambda$ is the cosmological constant. So what I have done in Eq.4.3, is replace the traditional cosmological term with $L_C$, hoping that it does a better job at representing non-local cosmological effects, than does $g^{½}\lambda$. The Lagrangian $L_G$ will certainly be free of Ostrogradsky instabilities if we take $L_C = L_{2C}$, but otherwise there maybe instability issues.

In the past (*see*, Bicknell and Klotz [29]) there was some interest in scalar-tensor theories that were invariant under a conformal transformation of the form

$$g_{ab} \rightarrow g'_{ab} := e^{2\sigma}g_{ab} \quad \text{and} \quad \varphi \rightarrow \varphi' := e^{-\sigma}\varphi .$$ Eq.4.4

The machinery that we developed here could be generalized to construct scalar-tensor field theories invariant under Eq.4.4. For such theories, flat space compatibility becomes the demand that $E^{ab}(L)$ and $E(L)$ be well-defined and differentiable when the metric tensor is flat and the scalar field vanishes. If $A^{ab} := E^{ab}(L)$, is $k^{th}$ order, then Aldersley's Identity for a theory invariant under Eq.4.4 in an n-dimensional space would be

$$\lambda^n A^{ab}( \, g; \, \partial g; \, \ldots ; \, \partial^k g; \, \varphi; \, \partial\varphi; \, \ldots ; \, \partial^k \varphi) =$$

$$= A^{ab}( \, g; \, \lambda\partial g; \, \ldots ; \, \lambda^k \, \partial^k g; \, \lambda\varphi; \, \lambda^2 \, \partial\varphi; \, \ldots ; \, \lambda^{k+1}\partial^k\varphi) \, . \qquad \text{Eq.4.5}$$

The identity for $B := E(L)$, is similar. Thus in a 4-dimensional space, $A^{ab}$ and $B$ can be at most of fourth-order in $g_{ab}$ and third-order in $\varphi$. One of the problems in building all $A^a_b$'s and $B$'s invariant under the transformation given in Eq.4.4, is that there is no simple generalization of Lemma 1 that enables us to spot admissible theories. I leave the task of building such theories to others.

This paper is intended to be the first of a series of three papers dealing with conformally invariant field theories. In the next paper I shall construct all conformally invariant, flat space compatible, vector-tensor field theories, that are consistent with charge conservation. The last paper will combine the results of the previous two papers to construct all conformally invariant, flat space compatible, scalar-vector-tensor field theories, that are consistent with charge conservation.

**Acknowledgements**

I wish to thank Dr.A.Guarnizo Trilleras, for making me aware of the subject of "beyond Horndeski Theories." He also introduced me to Dr.M.Zumalacárregui, who has provided me with great insight into numerous aspects of current scalar-tensor research, and assisted me in the analysis of Ostrogradsky instabilites.

I owe a debt of thanks to Professor J.M.Ezquiaga, for the interest he showed in the conformally invariant scalar-tensor theories that I presented in [11]. His remarks motivated me to construct the scalar-tensor theories that I presented here.

Lastly, I wish to thank my wife, Dr.S.Winklhofer Horndeski, for assistance in preparing this manuscript.

## Appendix A: Tensorial Concomitants and Their Derivatives

Not everyone is familiar with tensorial concomitants, or how to differentiate them with respect to their various arguments. The purpose of this appendix is to give the neophyte a crash course on these topics.

The quintessential tensorial concomitant is the curvature tensor. This concomitant associates to every pseudo-Riemannian manifold $V_n = (M,g)$, an $(0,4)$ tensor field R. If x is a chart of M with domain U, then the x-components of R are

$$R_{hijk} := g_{il} \left[ \Gamma^l_{hj,k} - \Gamma^l_{hk,j} - \Gamma^m_{hj} \Gamma^l_{mk} + \Gamma^m_{hk} \Gamma^l_{mj} \right], \qquad \text{Eq.A.1}$$

where $\Gamma^l_{hj}$ denotes the Christoffel symbols of the second kind. The important thing to note here, is that we use the same functions of the metric tensor and its first two derivatives, to compute the components of $R_{hijk}$, for every chart of every manifold. That observation is the essence of a tensorial concomitant. The form of these functions can be extracted from Eq.A.1. To that end, let $\{x_{ab}\}$, $\{x_{abc}\}$ and $\{x_{abcd}\}$ denote the standard coordinates of $\mathbb{R}^{nxn}$, $\mathbb{R}^{nxnxn}$ and $\mathbb{R}^{nxnxnxn}$ respectively. We now construct functions $\rho_{hijk} : \mathbb{R}^N \to \mathbb{R}$, where $N := n^2xn^3xn^4$, using Eq.A.1 as follows:

1) wherever you see $g_{ab}$ in $R_{hijk}$, you replace it by $\frac{1}{2}(x_{ab} + x_{ba})$;

2) wherever you see $g_{ab,c}$ in $R_{hijk}$, you replace it by $\frac{1}{2}(x_{abc} + x_{bac})$; and

3) wherever you see $g_{ab,cd}$ in $R_{hijk}$, you replace it by $\frac{1}{4}(x_{abcd}+x_{bacd}+x_{abdc}+x_{badc}) = x_{(ab)(cd)}$.

If P is any point in the domain of a chart x, then it is clear that $\rho_{hijk}(g_{ab}(P); g_{ab,c}(P); g_{ab,cd}(P)) = R_{hijk}(P)$. The real valued functions $\rho_{hijk}$ are the "absolute functions" that define $R_{hijk}$ on the domain of any chart of M. Note that these functions are differentiable, in fact real analytic.

At this point I can state the formal definition of a tensorial concomitant without turning people away in fear. We shall say that T is a $k^{th}$ order tensorial concomitant of a pseudo-Riemannian metric tensor of type (r,s,w), if T associates to each pseudo-Riemannian space $V_n = (M,g)$ a differentiable tensor field T(g) of type (r,s,w), and there exist real valued differentiable funtions $\tau^{i\cdots}_{j\ldots} : \mathbb{R}^N \to \mathbb{R}$ ($N := n^2xn^3x\ldots xn^{2+k}$), which are such that given any chart x of M with domain U, then the x components of T(g) on U are given by

$$T(g)^{i\cdots}{}_{j\ldots} = \tau^{i\cdots}{}_{j\ldots}(g_{ab}; g_{ab,c}; \ldots).$$

Evidently the Riemann curvature tensor is a second-order tensorial concomitant of a pseudo-Riemannian metric tensor of type (0,4,0).

Now that I have defined tensorial concomitants of a pseudo-Riemannian metric tensor, we shall turn our attention to differentiating such entities with respect to their various arguments. To illustrate how that is done let us consider the derivative of $R_{hijk}$ with respect to $g_{ab,cd}$. At first this quantity seems to be absurd, since we only know how to differentiate with respect to local coordinates of a manifold, and the $g_{ab,cd}$'s are not local coordinates. To circumvent that problem we define

$$\frac{\partial R_{hijk}}{\partial g_{ab,cd}} := \frac{\partial \rho_{hijk}}{\partial x_{abcd}}(g_{rs}; g_{rs,t}; g_{rs,tu}) \qquad \text{Eq.A.2}$$

The derivatives $\dfrac{\partial R_{hijk}}{\partial g_{ab}}$ and $\dfrac{\partial R_{hijk}}{\partial g_{ab,c}}$ are defined similarly. You should note that since $\rho_{hijk}$ was built to have the symmetries of $g_{ab}$, $g_{ab,c}$ and $g_{ab,cd}$, the derivatives with respect to $g_{ab}$, $g_{ab,c}$ and $g_{ab,cd}$ will have those symmetries. So that, *e.g.,*

$$\frac{\partial R_{hijk}}{\partial g_{ab,cd}} = \frac{\partial R_{hijk}}{\partial g_{ba,cd}} = \frac{\partial R_{hijk}}{\partial g_{ab,dc}} \qquad .$$

Now that you know how to formally define things like $\dfrac{\partial R_{hijk}}{\partial g_{ab,cd}}$, you never again use the standard coordinates of $\mathbb{R}^{n \times n}$ x $\mathbb{R}^{n \times n \times n}$ x $\mathbb{R}^{n \times n \times n \times n}$ to compute derivatives. You simply formally differentiate with respect to $g_{ab,cd}$ (say), with the understanding that wherever $g_{rs,tu}$ appears in the concomitant that you are differentiating, it has been replaced by $g_{(rs),(tu)}$. You then formally differentiate the resulting concomitant with respect to $g_{ab,cd}$ as if it had no symmetries. As a result we find that

$$\frac{\partial R_{hijk}}{\partial g_{ab,cd}} = \frac{\partial}{\partial g_{ab,cd}} [\tfrac{1}{2}(g_{hk,ij} + g_{ij,hk} - g_{hj,ik} - g_{ik,hj}) + \text{(lower order terms)}] =$$

$$= \tfrac{1}{8}[(\delta^a_h \, \delta^b_k + \delta^a_k \, \delta^b_h)(\delta^c_i \, \delta^d_j + \delta^c_j \, \delta^d_i) + (\delta^a_i \, \delta^b_j + \delta^a_j \, \delta^b_i)(\delta^c_h \, \delta^d_k + \delta^c_k \, \delta^d_h) +$$

$$- \; (\delta^a_h \, \delta^b_j + \delta^a_j \, \delta^b_h)(\delta^c_i \, \delta^d_k + \delta^c_k \, \delta^d_i) + (\delta^a_i \, \delta^b_k + \delta^a_k \, \delta^b_i)(\delta^c_h \, \delta^d_j + \delta^c_j \, \delta^d_h)] \; .$$

This agrees with the formula presented on page 313 of [13].

When dealing with tensorial concomitants involving a scalar field, $\varphi$, and its derivatives, the rules for differentiating with respect to $\varphi$, are similar to those for differentiating with respect to the metric tensor. *E.g.,* when differentiating with respect to $\varphi_{,abc}$ , you simply formally differentiate with respect to $\varphi_{,abc}$ , with the understanding that wherever $\varphi_{,rst}$ appears in the concomitant in question, it has been replaced by $\varphi_{,(rst)}$ . You then formally differentiate with respect to $\varphi_{,abc}$ as if it had no symmetries.

Armed with the knowledge of how to differentiate tensorial concomitants, you should be able to use Eqs.1.8-1.13 to compute Euler-Lagrange tensor densities of scalar-tensor Lagrangians. Two helpful things you need to know, and should derive, are that

$$\frac{\partial g}{\partial g_{hk}} = g g^{hk} \quad \text{and} \quad \frac{\partial g^{ij}}{\partial g_{hk}} = -\tfrac{1}{2}[g^{ih} \, g^{jk} + g^{ik} \, g^{hj}] \; .$$

Sometimes our definition of differentiation of tensorial concomitants gives rise to formulas that look wrong, but really aren't . To illustrate my point, consider the trivial concomitant $M_{ij} = M_{ij}(g_{rs}) := g_{ij}$ . According to our definition

$$\frac{\partial M_{ij}}{\partial g_{hk}} = \tfrac{1}{2}(\delta^h_i \, \delta^k_j + \delta^h_j \, \delta^k_i) \ .$$

So if we let h=i=1, and j=k=2, we get

$$\frac{\partial M_{12}}{\partial g_{12}} = \frac{\partial g_{12}}{\partial g_{12}} = \tfrac{1}{2} \ ,$$

which seems ridiculous. It seems crazy because, at first you think $g_{12}$ is like the x coordinate in $\mathbb{R}^4$, and $\partial x / \partial x = 1$. But $g_{12}$ is not like a coordinate. The $g_{ij}$'s are functions of the four local coordinates, and the symbol $\partial g_{12}/\partial g_{12}$, requires a definition, before you can proceed to compute it. I have provided that definition, and our differentiation process is a solid as calculus in $\mathbb{R}^n$. So there is nothing to be concerned about, so long as you understand second year calculus, and follow its rules.

Let me add that there exists another completely consistent, and more "modern" way, to define tensorial concomitants and their derivatives. This involves using what are called "jet bundles." This is kind of similar to what we have done, but requires a lot more work before you can introduce coordinates, and then the coordinates for second-order concomitants of the metric tensor are $\{x_{ab}\}, \{x_{abc}\}$ and $\{x_{abcd}\}$, restricted so that $a \leq b$ in $x_{ab}$ , $a \leq b$ in $x_{abc}$, and $a \leq b$, $c \leq d$ in $x_{abcd}$. As you might expect, the formulas for $\rho_{hijk}$ and its derivatives in this situation are quite a mess, since you can no longer use the summation convention, and must do sums that respect the ranges of $x_{ab}$, $x_{abc}$ and $x_{abcd}$.

The approach that I have described here, for differentiating tensorial concomitants, is much easier than the jet bundle approach, and is the one followed by Lovelock and Rund in [13].

## Appendix B: Thomas's Tensor Extensions

Let $V_n = (M, g)$ be an n-dimensional pseudo-Riemannian space, and let T be a (1,1) tensor field in M. If P is a point in the domain of T, and x is a chart at P, we define the x components of the first, second, third,... extensions of T at P by

$$T^i_{j;k}(P) := \frac{\partial t^i_j}{\partial y^k}(P) \, , \, T^i_{j;kl}(P) := \frac{\partial^2 t^i_j}{\partial y^k \partial y^l}(P) \, , \, T^i_{j;klm}(P) := \frac{\partial^3 t^i_j}{\partial y^k \partial y^l \partial y^m}(P) \, ,... \qquad \text{Eq.B.1}$$

where $t^i_j$ denotes the components of T with respect to the normal coordinate system y at P determined so that $\partial/_{\partial x}{}^i = \partial/_{\partial yi}$ at P. The extensions of T are tensor fields. I shall explain why this is so for the first extension. The proof for the higher extensions is similar. Let x' be another chart at P, with corresponding normal coordinate system y'. We need to prove that at P

$$T'^i_{j;k} = T^a_{b;c} \frac{\partial x'^i}{\partial x^a} \frac{\partial x^b}{\partial x'^j} \frac{\partial x^c}{\partial x'^k} \, , \qquad \text{Eq.B.2}$$

where at P

$$T'^i_{j;k} = \frac{\partial t'^i_j}{\partial y'^k} \quad .$$ 

<div align="right">Eq.B.3</div>

Since T is a (1,1) tensor

$$t'^i_j = t^a_b \frac{\partial y'^i}{\partial y^a} \frac{\partial y^b}{\partial y'^j} \quad ,$$

and so at P

$$T'^i_{j;k} = \frac{\partial t^a_b}{\partial y^c} \frac{\partial y^c}{\partial y'^k} \frac{\partial y'^i}{\partial y^a} \frac{\partial y^b}{\partial y'^j} + t^a_b \frac{\partial}{\partial y'^k} \left[ \frac{\partial y'^i}{\partial y^a} \frac{\partial y^b}{\partial y'^j} \right] \quad .$$

<div align="right">Eq.B.4</div>

To simplify this expression we need to know the relationship between the two charts, y and y'. To that end let $v \in T_pM$, and let $c_v = c_v(s)$, denote the affinely parameterized geodesic emanating from P with initial tangent vector v. Due to the definition of normal coordinates we have

$$y^i \circ c_v(s) = v^i s \text{ , and } y'^i \circ c_v(s) = v'^i s$$

<div align="right">Eq.B.5</div>

where, due to the vector transformation law, we have

$$v^i = v'^j \frac{\partial x^i}{\partial x'^j}\bigg|_p \quad .$$

<div align="right">Eq.B.6</div>

Since each point Q in a neighborhood of P is expressible as $c_v(1)$ for some $v \in T_pM$, we can use Eq.B.5 to deduce that $y^i(Q) = v^i$ and $y'^i(Q) = v'^i$. Thus due to Eq.B.6 we know that for every Q in a neighborhood of P, $y^i(Q) = y'^j(Q) \frac{\partial x^i}{\partial x'^j}\bigg|_p$ . Hence if y and y' are normal coordinates at P, then on a neighborhood of P

$$y^i = y'^j \frac{\partial x^i}{\partial x'^j}\bigg|_p \quad ,$$

<div align="right">Eq.B.7</div>

and so they are linearly related.

We can combine Eqs.B.4 and B.7 to deduce that Eq.B.2 is indeed valid, and so the first extension of the (1,1) tensor T is a (1,2) tensor field in M. In a similar way we can prove that all of the higher extensions of T are tensor fields. This explains why I said that all of the derivatives of $\gamma_{ij}$ and $\varphi$ on the right-hand side of Eq.2.22, are tensors.

Not everyone has access to Thomas's work [24]. So I thought that I would present Thomas's arguments, in my own words, to show how the extensions of a tensor can be represented in terms of quantities that we are more familiar with, like the Christoffel symbols, and their derivatives. To show how this can be done lets consider the third extension of $\varphi$.

If x is chart at P, and y is its corresponding normal coordinate system at P, then

$$\varphi_{:ijk} := \frac{\partial^3 \varphi}{\partial y^i \partial y^j \partial y^k}\bigg|_p \ .$$

Eq.B.8

Using the rule for differentiating composite functions, and the symmetries of $\varphi_{,ab}$ , we find that

$$\frac{\partial \varphi}{\partial y^i} = \frac{\partial \varphi}{\partial x^a}\frac{\partial x^a}{\partial y^i} \ , \quad \frac{\partial^2 \varphi}{\partial y^i \partial y^j} = \frac{\partial^2 \varphi}{\partial x^a \partial x^b}\frac{\partial x^a}{\partial y^i}\frac{\partial x^b}{\partial y^j} + \frac{\partial \varphi}{\partial x^a}\frac{\partial^2 x^a}{\partial y^i \partial y^j} \ \text{ and}$$

$$\frac{\partial^3 \varphi}{\partial y^i \partial y^j \partial y^k} = \frac{\partial^3 \varphi}{\partial x^a \partial x^b \partial x^c}\frac{\partial x^a}{\partial y^i}\frac{\partial x^b}{\partial y^j}\frac{\partial x^c}{\partial y^k} + \frac{\partial^2 \varphi}{\partial x^a \partial x^b}\bigg[\frac{\partial^2 x^a}{\partial y^i \partial y^j}\frac{\partial x^b}{\partial y^k} + \frac{\partial^2 x^a}{\partial y^i \partial y^k}\frac{\partial x^b}{\partial y^i} \ +$$

$$+ \frac{\partial^2 x^a}{\partial y^k \partial y^i} \frac{\partial x^b}{\partial y^j} \Bigg] \;\; + \;\; \frac{\partial \varphi}{\partial x^a} \frac{\partial^3 x^a}{\partial y^i \partial y^j \partial y^k} \quad . \qquad\qquad \text{Eq.B.9}$$

The problem now is, how does one compute the partial derivatives of $x^a$ with respect to the $y^i$'s?  To determine these quantities we shall exploit the geodesic equations, which are a veritable goldmine of information.

Let P be a point in our manifold, and let x be a chart at P with corresponding normal chart y.  If $v \in T_P M$, and $c_v(s)$ is the corresponding affinely parameterized geodesic emanating from P, we set  $x^a(s) := x^a \circ c_v(s)$, and $y^a(s) := y^a \circ c_v(s) = v^a s$, due to Eq.B.5.  Thus we have

$$\frac{dx^a(s)}{ds} = \frac{\partial x^a}{\partial y^i} \frac{dy^i}{ds} = \frac{\partial x^a}{\partial y^i} v^i \; ,$$

and hence for every k=1,2,... we can show that

$$\frac{d^k x^a(s)}{ds^k} = \frac{\partial^k x^a}{\partial y^{i_1} ... \partial y^{i_k}} v^{i_1} ... v^{i_k} \; . \qquad\qquad \text{Eq.B.10}$$

So we now see that if we can just find expressions for the derivatives of $x^i(s)$ with respect to s, then we can use Eq.B.10 to help us simplify our formula for $\varphi_{;ijk}$ given in Eq.B.9.  The geodesic equation will come to our rescue here.

The equation for affinely parameterized geodesics is

$$\frac{d^2 x^a}{ds^2} + \Gamma^a_{ij} \frac{dx^i}{ds} \frac{dx^j}{ds} \;\; = 0, \qquad\qquad \text{Eq.B.11}$$

where, following custom, I have let $x^a = x^a(s)$ in this  equation.   If  we  differentiate

Eq.B.11 with respect to s we get

$$\frac{d^3x^a}{ds^3} + \Gamma^a_{ij,k}\frac{dx^i}{ds}\frac{dx^j}{ds}\frac{dx^k}{ds} + \Gamma^a_{ij}\frac{d^2x^i}{ds^2}\frac{dx^j}{ds} + \Gamma^a_{ij}\frac{dx^i}{ds}\frac{d^2x^j}{ds^2} = 0 \ . \qquad\text{Eq.B.12}$$

Using Eq.B.11 in Eq.B.12 to eliminate the second derivatives we discover that

$$\frac{d^3x^a}{ds^3} + \Gamma^a_{ijk}\frac{dx^i}{ds}\frac{dx^j}{ds}\frac{dx^k}{ds} = 0 \ , \qquad\qquad\text{Eq.B.13}$$

where

$$\Gamma^a_{ijk} := \Gamma^a_{(ij,k)} - 2\,\Gamma^a_{m(i}\,\Gamma^m_{jk)} \ . \qquad\qquad\text{Eq.B.14}$$

Repeating this argument once again we find that

$$\frac{d^4x^a}{ds^4} + \Gamma^a_{ijkl}\frac{dx^i}{ds}\frac{dx^j}{ds}\frac{dx^k}{ds}\frac{dx^l}{ds} = 0 \ , \qquad\qquad\text{Eq.B.15}$$

where

$$\Gamma^a_{ijkl} := \Gamma^a_{(ijk,l)} - 3\Gamma^a_{m(ij}\,\Gamma^m_{kl)} \ . \qquad\qquad\text{Eq.B.16}$$

There are similar formulas for $\frac{d^k x^i}{ds^k}$ , but we shall not require them in what follows.
We can now employ Eqs.B.10, B.11, B.13 and B.15 to show that at P

$$\frac{\partial^2 x^a}{\partial y^i\partial y^j} = -\Gamma^a_{ij}\ , \quad \frac{\partial^3 x^a}{\partial y^i\partial y^j\partial y^k} = -\Gamma^a_{ijk}\ , \text{ and } \frac{\partial^4 x^i}{\partial y^i\partial y^j\partial^k y\partial y^l} = -\Gamma^a_{ijkl}\ . \qquad\text{Eq.B.17}$$

Using Eqs.B.8, and B.17, we find that Eq.B.9 implies $\varphi_{;ijk} = \varphi_{(ijk)}$ . In a similar way, a lengthy calculation shows that $\varphi_{;ijkl} = \varphi_{(ijkl)}$ .

Thomas never gives a name to the quantities $\Gamma^a_{ijk}, \Gamma^a_{ijkl}$ , appearing in Eqs.B.14 and B.16, nor to their higher order counterparts, which arise in the k[th] derivative of $x^a$ with respect to s. I suggest that we just refer to them as the higher order Christoffel symbols of the second kind. One should note that these quantities are completely symmetric in their covariant indices.

Two other quantities we need for Lemma 4 are the "metric normal tensors," $g_{ij;kl}$ and $g_{ij;klm}$. Expressions for these entities with respect to an arbitrary chart x can be found using Eqs.33.5-33.7 on page 99 of Thomas's book [24]. (Note that Thomas denotes extensions with a comma, while I am using a semi-colon.) The trick to evaluating Thomas's formulas (as well as Eq.B.9 above), is to evaluate them at the pole of a normal coordinate system where $\Gamma^a_{bc} = 0$. In that case Thomas's formulas for $g_{ij;kl}$ and $g_{ij;klm}$ become

$$g_{ij;kl} = g_{ij,kl} - g_{aj}\,\Gamma^a_{ikl} - g_{ia}\,\Gamma^a_{jkl} \qquad\qquad \text{Eq.B.18}$$

$$g_{ij;klm} = g_{ij,klm} - g_{aj}\,\Gamma^a_{iklm} - g_{ia}\,\Gamma^a_{jklm} \qquad\qquad \text{Eq.B.19}$$

where at the pole

$$\Gamma^a_{jkl} = \tfrac{1}{3}(\Gamma^a_{jk,l} + \Gamma^a_{kl,j} + \Gamma^a_{lj,k}) \qquad\qquad \text{Eq.B.20}$$

$$\Gamma^a_{jklm} = \tfrac{1}{4}(\Gamma^a_{jkl,m} + \Gamma^a_{klm,j} + \Gamma^a_{lmj,k} + \Gamma^a_{mjk,l}) \qquad\qquad \text{Eq.B.21}$$

and

$$\Gamma^a_{jkl,m} = \tfrac{1}{3}(\Gamma^a_{jk,lm} + \Gamma^a_{kl,jm} + \Gamma^a_{lj,km})\,. \qquad\qquad \text{Eq.B.22}$$

Upon combining Eqs.B.18-B.22, we find that at the pole of a geodesic coordinate system

$$g_{ij;kl} = \tfrac{1}{3}(R_{iklj} + R_{ilkj}) \qquad\qquad \text{Eq.B.23}$$

and

$$g_{ij;klm} = \tfrac{1}{6}(R_{iklj|m} + R_{ilkj|m} + R_{ilmj|k} + R_{imlj|k} + R_{imkj|l} + R_{ikmj|l})\,. \qquad\qquad \text{Eq.B.24}$$

Since the above equations are manifestly tensorial they hold at all points of our manifold.

If you wish to derive the general formulas for $g_{ij;kl}$ and $g_{ij;klm}$ *ab initio*, you can do so by starting with the definition of these quantities which gives us

$$g_{ij;kl} = \frac{\partial^2 \gamma_{ij}}{\partial y^k \partial y^l} \quad \text{and} \quad g_{ij;klm} = \frac{\partial^3 \gamma_{ij}}{\partial y^k \partial y^l \partial y^m}$$

at a fixed point P, with

$$\gamma_{ij} = g_{ab} \frac{\partial x^a}{\partial y^i} \frac{\partial x^b}{\partial y^j}.$$

To simplify the ensuing expressions use Eq.B.17.

In concluding this section I would like to point out that the first extension of a tensor is the same as its covariant derivative. However, this is not the case for the higher extensions. For example, due to Eq.B.23, $g_{ij;kl} \neq g_{ij|kl} = g_{ij;k;l} = 0$.

## Appendix C: Constructing Tensorial Concomitants of $g_{ab}$ and $\varphi$

There are essentially two approaches to the construction of tensorial concomitants of the metric tensor and scalar field--the differential approach, and the algebraic approach. I was "raised" on the differential approach, which was developed by my "mathematical forefathers" professors Rund and Lovelock. This approach is illustrated on pages 312-317 of their book [13]. As we shall see, the differential approach is also very algebraic in nature. The algebraic approach was introduced to me by my student Dr.Aldersley, who used it in [22] and [23]. I believe that Hilbert developed the algebraic approach, and it is described in detail in Weyl's book [30].

Rather than discuss the two approaches in general terms, let's consider the problem of constructing the tensor density concomitant $\Phi^{abcdef}$ of $g_{ij}$ and $\varphi$, which has the following symmetries:

$$\Phi^{abcdef} = \Phi^{(ab)cdef} = \Phi^{ab(cde)f} , \; g_{ab}\Phi^{abcdef} = 0, \text{ and } \Phi^{a(bcde)f} = 0. \qquad \text{Eq.C.1}$$

We need $\Phi^{abcdef}$ in Lemma 6. In both the differential and algebraic approaches it is simpler to deal with the covariant tensor version of $\Phi^{abcdef}$, and not the contravariant tensor density. This tensor will be denoted by $\Phi_{abcdef}$, and should not cause any confusion.

I shall now outline how the differential approach can be used to build $\Phi_{abcdef}$. Let P be any point in our manifold, and let x and x' be two charts at P. Due to the tensoriality of $\Phi_{abcdef}$ we can write

$$\Phi_{abcdef}(g'_{rs}, \varphi') = \Phi_{hijklm}(g_{rs}, \varphi)J_a{}^h J_b{}^i J_c{}^j J_d{}^k J_e{}^l J_f{}^m \qquad \text{Eq.C.2}$$

where

$$g'_{rs} = g_{pq}J_r{}^p J_s{}^q , \; \varphi' = \varphi \quad J_a{}^h = \frac{\partial x^h}{\partial x'^a} .$$

Upon differentiating Eq.C.2 with respect to $J_v{}^u$, and then evaluating the result for the identity transformation, we obtain

$$2\Phi_{abcdef}{}^{;vs}g_{us} =$$

$$= \Phi_{ubcdef}\delta^{va} + \Phi_{aucdef}\,\delta^v{}_b + \Phi_{abudef}\,\delta^v{}_c + \Phi_{abcuef}\,\delta^v{}_d + \Phi_{abcduf}\,\delta^v{}_e + \Phi_{abcdeu}\,\delta^v{}_f , \qquad \text{Eq.C.3}$$

where

$$\Phi_{abcdef}{}^{:rs} = \frac{\partial \Phi_{abcdef}}{\partial g_{rs}} \quad .$$

This completes all of the differentiation there is in the differential approach to building concomitants of $g_{ab}$ and $\varphi$! Now for a lot of algebra.

If we multiply Eq.C.3 by $g^{uw}$, we find that the left-hand side of the resulting equation is symmetric in $v$ and $w$. This implies that the right-hand side must also be symmetric in $v$ and $w$. Consequently we must have

$$\Phi_{ubcdef}\, g^{uw}\, \delta^v{}_a + \Phi_{aucdef}\, g^{uw}\delta^v{}_b + \Phi_{abudef}\, g^{uw}\delta^v{}_c + \Phi_{abcuef}\, g^{uw}\delta^v{}_d + \Phi_{abcduf}\, g^{uw}\delta^v{}_e + \Phi_{abcdeu}g^{uw}\delta^v{}_f =$$

$$\Phi_{ubcdef}\, g^{uv}\, \delta^w{}_a + \Phi_{aucdef}\, g^{uv}\delta^w{}_b + \Phi_{abudef}\, g^{uv}\delta^w{}_c + \Phi_{abcuef}\, g^{uv}\delta^w{}_d + \Phi_{abcduf}\, g^{uv}\delta^w{}_e + \Phi_{abcdeu}\, g^{uv}\delta^w{}_f \,.$$

Upon contracting the above equation on $v$ and $a$, and then multiplying the result by $g_{wa}$ we obtain (where, for the moment, we are assuming that we are working in an n-dimensional space)

$$n\Phi_{abcdef} + \Phi_{bacdef} + \Phi_{cbadef} + \Phi_{dbcaef} + \Phi_{ebcdaf} + \Phi_{fbcdea} = \Phi_{abcdef} + \Phi_{vucdef}g^{uv}g_{ab} + \Phi_{ubvdef}g^{uv}g_{ac} +$$

$$+\Phi_{vbcuef}g^{uv}g_{ad} + \Phi_{vbcduf}\, g^{uv}g_{ae} + \Phi_{vbcdeu}\, g^{uv}\, g_{af} \,. \hspace{3cm} \text{Eq.C.4}$$

This equation can be rewritten as

$$(n-1)\Phi_{abcdef} + \Phi_{bacdef} + \Phi_{cbadef} + \Phi_{dbcaef} + \Phi_{ebcdaf} + \Phi_{fbcdea} = \Xi_{abcdef} \,, \hspace{1.5cm} \text{Eq.C.5}$$

where

$$\Xi_{abcdef} := \Phi_{vucdef}\, g^{uv}g_{ab} + \Phi_{ubvdef}\, g^{uv}g_{ac} + \Phi_{vbcuef}\, g^{uv}g_{ad} + \Phi_{vbcduf}g^{uv}\, g_{ae} + \Phi_{vbcdeu}g^{uv}g_{af}\,. \hspace{0.3cm} \text{Eq.C.6}$$

Owing to the symmetries enjoyed by $\Phi_{abcdef}$, as listed in Eq.C.1, the second through fifth terms on the right-hand side of Eq.C.5 add up to zero, and the first term in $\Xi_{abcdef}$ vanishes. Hence Eq.C.5 becomes

$$(n-1)\Phi_{abcdef} + \Phi_{fbcdea} = \Xi_{abcdef} \ . \qquad\qquad \text{Eq.C.7}$$

Upon interchanging the indices a and f in Eq.C.7 we obtain

$$(n-1)\Phi_{fbcdea} + \Phi_{abcdef} = \Xi_{fbcdea} \ . \qquad\qquad \text{Eq.C.8}$$

Eqs.C.7 and C.8 combine to tell us that if $n \neq 2$ then

$$\Phi_{abcdef} = (n(n-2))^{-1}((n-1)\Xi_{abcdef} - \Xi_{fbcdea}) \ . \qquad\qquad \text{Eq.C.9}$$

Thus we see that we have expressed the (0,6) tensor $\Phi_{abcdef}$ in terms of (0,4) tensors in $\Xi_{abcdef}$. These (0,4) tensors have various symmetries due to Eq.C.1, and as a result can be constructed from (0,2) tensors using the (0,4) version of Eq.C.4. In a 4-dimensional space, the (0,2) tensors must just be scalar functions of $\varphi$ times $g_{ab}$. ( The proof of that elementary fact will be given below.) The important thing to note is that in a 4-dimensional space $\Phi_{abcdef}$ will turn out to be a product of three $g_{ab}$'s with coefficients that are scalar functions of $\varphi$.

At this point you might ask, what would have happened if $\Phi_{abcdef}$ did not have all of those symmetries which allowed us to replace Eq.C.5 by Eq.C.7? In that case what you do is this (and in most cases you can get away with less): Consider the linear system of equations for the components of $\Phi_{abcdef}$ obtained by considering all 6! permutations of a,b,c,d,e,f in Eq.C.5. This will give rise to a linear system of 6!

equations for 6! variables. You then solve that system for $\Phi_{abcdef}$ in terms of linear combinations of the $\Xi_{abcdef}$'s, which in turn are built from the $g_{ab}$'s and (0,4) tensorial conconmitants of $g_{ab}$ and $\varphi$. The (0,4) concomitants are then built using a technique similar to the one just employed to build $\Phi_{abcdef}$. In this way $\Phi_{abcdef}$ can be assembled. The problem with this approach is whether the 6!x6! matrix that generated our linear system of 6! equations for $\Phi_{abcdef}$ is invertible. If it isn't, then the system of equations needs to be augmented with additional equations. From my experience it is always clear how to proceed. Perhaps a simple example is in order.

Say we want to build $\Phi_{ab} = \Phi_{ab}(g_{rs}, \varphi)$. Then the obvious counterpart of Eq.C.5 is

$$(n-1)\Phi_{ab} + \Phi_{ba} = \Xi_{ab} \qquad\qquad \text{Eq.C.10}$$

where

$$\Xi_{ab} = \Phi_{uv}g^{uv}\,g_{ab}\,.$$

Assume $\Phi_{ab}$ has no symmetries. Upon permuting a and b in Eq.C.10 we get

$$(n-1)\Phi_{ba} + \Phi_{ab} = \Xi_{ba}\,. \qquad\qquad \text{Eq.C.11}$$

Eqs.C.10 and C.11 can be solved for $\Phi_{ab}$ provided $n \neq 2$. When n=2, the system of 2! equations generated by Eq.C.10 is degenerate, and needs to be supplemented in order to determine $\Phi_{ab}$. Well, in a 2-dimensional space, the anti-symmetric part of $\Phi_{ab}$ has only one component. Thus we may write $\Phi_{[ab]} = \alpha(\varphi)e_{ab}$, where $e_{ab}$ is the Levi-Civita tensor in a 2-dimensional space, and $\alpha$ is an arbitrary scalar function of $\varphi$. Since $\Phi_{ab}=$

$\Phi_{(ab)} + \Phi_{[ab]}$ , we can use Eq.C.10 to write

$$\Phi_{ab} = \alpha(\varphi)e_{ab} + \beta(\varphi)g_{ab}$$

in a 2-dimensional space, where $\beta$ is a scalar function of $\varphi$.

The problem we encountered when trying to build $\Phi_{ab}$ in a 2-dimensional space, also happens when you try to build an $(0,k)$ tensor concomitant of $g_{rs}$ and $\varphi$, in a space of dimension n=k. In that case the linear system of n! equations will need to be supplemented with the equation $\Phi_{[a...k]} = \alpha(\varphi)e_{a...k}$, where $e_{....}$ is the Levi-Civita tensor. I think that is all we need to do to guarantee that the system of k! equations will give us a unique solution for $\Phi_{a...k}$ in terms of $(0,k-2)$ tensorial concomitants of $g_{rs}$ and $\varphi$. However, I do not have a rigorous proof that this is so.  All I can say is that for every example I have encountered, that is how it always worked out.

The upshot of the above analysis is that the differential approach can provide quite an arduous path to determining $(0,k)$ tensor concomitants of $g_{ab}$ and $\varphi$.  That is why the algebraic approach, which I shall now describe, is so useful.

From the modern definition of $(0,k)$ tensors, we know that at each point P, of a manifold M, they give rise to a multilinear form mapping $(T_PM)^k$ into $\mathbb{R}$. If $\Phi$ is an $(0,k)$ tensor field, then its associated map is: $\Phi(v_1,...,v_k) := \Phi_{a(1)...a(k)} v_1^{a(1)}...v_k^{a(k)}$, where for every $\alpha = 1,...,k,$ $v_\alpha = v_\alpha{}^a \dfrac{\partial}{\partial x^a}\Big|_P$, and x is a chart at P.  In Weyl's book *(see,* page 23 of  [30]), whenever he speaks of a multilinear form on a vector space, he has

something like $\Phi(v_1,...,v_k)$ in mind.  From the results Weyl presents in sections 8 and 12 of [30], we can conclude that in a 4-dimensional space, an arbitrary (0,k) tensorial concomitant of $g_{ab}$ and $\varphi$, is generated from all possible products of $g_{ab}$ and $e_{abcd}$ (the 4-dimensional Levi-Civita tensor), with coefficients which are scalar functions of $\varphi$. That is essentially what our differential approach to constructing such concomitants was trying to tell us; however, I could not supply all the details of the proof. Weyl does supply the details for his proof.

The important thing to note, as Weyl repeatedly mentions,  is that the "all possible products" of $g_{ab}$ and $e_{abcd}$ one uses to build (0,k) tensorial concomitants of $g_{ab}$ and $\varphi$, generates the space of tensors, but need not be a basis; *i.e.,* they may not form a linearly independent set.  *E.g.,* we have the identity $g_{a[b}e_{cdef]} = 0$ , and other such things to watch out for.

To illustrate the algebraic approach let us  build the tensor $\Phi_{abcdef}$ , whose symmetries are given by Eq.C.1. To begin, can there be any terms of the form $g_{..}e_{....}$ in $\Phi_{abcdef}$? Well, let us try putting indices into $e_{....}$. The indices in $\Phi_{abcdef}$ are grouped into three symmetric blocks: (ab), (cde), and f.  So if we put an "a" into $e_{....}$, then we can not put a "b" in $e_{....}$.  Likewise, if we put a "c" in $e_{....}$, then we can not put a "d" or and "e" into $e_{....}$. So we see that it is easy to put three indices from a,b,c,d,e,f  into $e_{....}$, but then we can not put a fourth index into $e_{....}$, which is compatible with $\Phi_{abcdef}$'s symmetries.

Thus $\Phi_{abcdef}$ must be built from a product of three $g$ 's.  So our initial expression for $\Phi_{abcdef}$ is

$$\Phi_{abcdef} = \qquad\qquad\qquad\qquad \text{Eq.C.12}$$

$\alpha_1(ab)(cd)(ef) + \; \alpha_2(ab)(ce)(fd) + \; \alpha_3(ab)(cf)(de) + \alpha_4(ac)(de)(fb) + \; \alpha_5(ac)(df)(be) +$

$\alpha_6(ac)(db)(ef) + \; \alpha_7(ad)(ef)(bc) + \; \alpha_8(ad)(eb)(cf) + \alpha_9(ad)(ec)(fb) + \alpha_{10}\,(ae)(fb)(cd) +$

$\alpha_{11}(ae)(fc)(db) + \alpha_{12}(ae)(fd)(bc) + \alpha_{13}(af)(bc)(de) + \alpha_{14}(af)(bd)(ec) + \alpha_{15}(af)(be)(cd),$

where the $\alpha$'s are scalar functions of $\varphi$, and $(ab) := g_{ab}$ .  Before we begin to determine the required relationships between the $\alpha$'s which guarantee that $\Phi_{abcdef}$ has the necessary symmetries, we need to demonstrate that the terms in Eq.C.12, forms a linearly independent set.  That will be the case if the equation $\Phi_{abcdef} = 0$, implies that all the $\alpha$'s must vanish.  To see that this is so, let $\{X_\beta,\; \beta{=}1\ \text{to}\ 4\}$, be an orthonormal set of vectors at an arbitrary point P of space.  Thus $g(X_\beta,X_\gamma) = \varepsilon_\beta \delta_{\beta\gamma}$ (no sum on $\beta$), with $\varepsilon_\beta = +1$, or $-1$.  Now it is easily seen that $\Phi(X_1, X_1, X_2, X_2, X_3, X_3) = \varepsilon_1 \varepsilon_2 \varepsilon_3 \alpha_1$, and thus $\Phi = 0$, implies $\alpha_1 = 0$.  In a similar way we can prove that $\Phi = 0$, implies that all the other $\alpha$'s must vanish.  Consequently the products of three $g$ 's appearing in Eq.C.12 forms a linearly independent set.  So we can now start imposing constraints upon the coefficients appearing in Eq.C.12.

Two of $\Phi_{abcdef}$'s symmetries are $\Phi_{abcdef} = \Phi_{(ab)cdef} = \Phi_{ab(cde)f}$ .Using Eq.C.12 we find that if $\Phi_{abcdef}$ satisfies these symmetries then it must have the following form

$$\Phi_{abcdef} = \alpha_1[(ab)(cd)(ef) + (ab)(ce)(fd) + (ab)(cf)(de)] + \qquad \text{Eq.C.13}$$

$$+\alpha_4[(ac)(bf)(de)+(af)(bc)(de)+(ad)(bf)(ce)+(af)(bd)(ce)+(ae)(bf)(cd)+(af)(be)(cd)]+$$

$$+\alpha_5[(ac)(be)(df)+(ae)(bc)(df)+(ad)(be)(cf)+(ae)(bd)(cf)+(ac)(bd)(ef)+(ad)(bc)(ef)].$$

If we now require that $g^{ab}\Phi_{abcdef} = 0$, we can use Eq.C.13 to deduce that

$$\alpha_1 = -\tfrac{1}{2}(\alpha_4 + \alpha_5) \; . \qquad\qquad \text{Eq.C.14}$$

The last condition that $\Phi_{abcdef}$ must satisfy is $\Phi_{a(bcde)f} = 0$. When this condition is imposed on the expression for $\Phi_{abcdef}$ presented in Eq.C.13, we can use Eq.C.14 to conclude that $\Phi_{abcdef}$ must vanish, as claimed in Lemma 6.

Now that we have finished our warm up exercise, we can tackle $\Psi_{abcdefhi}$. This tensor has the following symmetries:

$$\Psi_{abcdefhi} = \Psi_{(ab)cdefhi} = \Psi_{ab(cd)efhi} = \Psi_{abcd(efh)i} \; ; \; g^{ab}\,\Psi_{abcdefhi} = 0, \; g^{cd}\,\Psi_{abcdefhi} = 0, \quad \text{Eq.C.15}$$

$$\Psi_{a(b|cd|efh)i} = 0, \quad \text{and} \quad \Psi_{abc(defh)i} = 0 \; . \qquad\qquad \text{Eq.C.16}$$

Let V denote the space of tensors which satisfy Eqs.C.15 and C.16. V is a direct sum of three vector spaces, which are subspaces of the space of (0,8) tensors. We can express V as, $V = V_0 \oplus V_1 \oplus V_2$, where:

1) $V_0$ is built from the product of zero e's, and four g's;

2) $V_1$ is built from the product of one e, and two g's ; and

3) $V_2$ is built from the product two e's, and no g's.

In order to construct V, let us first look at $V_2$. How can we multiply two e's together to obtain something that satisfies Eq.C.15? Say we start with one e. It can not have two indices that appear in one of $\Phi_{abcdefhi}$ three blocks of symmetric indices. This is so since it would vanish when we symmetrize over those indices. So e$_{..}$ must have one index from a,b; one index from c,d; and one index from e,f,h; and then its fourth index would have to be i. Once these choices are made, there is no possible choice for the indices of the second e, that yields a term compatible with Eq.C.15. Therefore we have demonstrated that $V_2 = \{0\}$.

$V_1$ is another matter altogether. $V_1$ is not trivial, and our candidate for a general element if $V_1$ is:

$$\Psi_{abcdefhi} =$$

$\alpha_1(acei)(bd)(fh)+\alpha_2(acei)(bf)(hd)+\alpha_3(acei)(bh)(df)+\alpha_4(acfi)(bd)(eh)+\alpha_5(acfi)(be)(hd)+$
$\alpha_6(acfi)(bh)(de)+\alpha_7(achi)(bd)(ef)+\alpha_8(achi)(be)(fd)+\alpha_9(achi)(bf)(de)+\alpha_{10}(adei)(bc)(fh)+$
$\alpha_{11}(adei)(bf)(hc)+\alpha_{12}(adei)(bh)(cf)+\alpha_{13}(adfi)(bc)(eh)+\alpha_{14}(adfi)(be)(hc)+\alpha_{15}(adfi)(bh)(ce)+$
$\alpha_{16}(adhi)(bc)(ef)+\alpha_{17}(adhi)(be)(fc)+\alpha_{18}(adhi)(bf)(ce)+\alpha_{19}(bcei)(ad)(fh)+\alpha_{20}(bcei)(af)(hd)+$
$\alpha_{21}(bcei)(ah)(df)+\alpha_{22}(bcfi)(ad)(eh)+\alpha_{23}(bcfi)(ae)(hd)+\alpha_{24}(bcfi)(ah)(de)+\alpha_{25}(bchi)(ad)(ef)+$
$\alpha_{26}(bchi)(ae)(fd)+\alpha_{27}(bchi)(af)(de)+\alpha_{28}(bdei)(ac)(fh)+\alpha_{29}(bdei)(af)(hc)+\alpha_{30}(bdei)(ah)(cf)+$
$\alpha_{31}(bdfi)(ac)(eh)+\alpha_{32}(bdfi)(ae)(hc)+\alpha_{33}(bdfi)(ah)(ce)+\alpha_{34}(bdhi)(ac)(ef)+\alpha_{35}(bdhi)(ae)(fc)+$
$\alpha_{36}(bdhi)(af)(ce)$ ,                                        Eq.C.17

where $(acei) := e_{acei}$ , $(bd):= g_{bd}$, and $\alpha_1,\ldots,\alpha_{36}$ are scalar functions of $\varphi$. Before we can examine the consequences that Eqs.C.15 and C.16 have for the coefficients of our initial expression for an element of $V_1$, we need to demonstrate that the products of two g's and one e in Eq.C.17 form a linearly independent set. Once again this is easy to prove. Let P be any point of our manifold M, and let $\{X_\alpha; \alpha=1,...,4\}$ be an orthonormal basis for $T_PM$. We need to show that $\Psi_{abcdefhi} = 0$ in Eq.C.17, implies that all of the $\alpha$'s must vanish. To achieve that end we note that if we multiply the equation $\Psi_{abcdefhi} = 0$, by $X_4^{\ b} X_4^{\ i}$ we obtain an equation which only involves $\alpha_{1},...,\alpha_{18}$. Similarly, if we multiply the equation $\Psi_{abcdefhi} = 0$, by $X_4^{\ a} X_4^{\ i}$ we obtain an equation that only involves $\alpha_{19},...,\alpha_{36}$. Now what we want are combinations of eight X's, which are such that when they are multiplied times the equation $\Psi_{abcdefhi} = 0$, they pick out any $\alpha$ we wish. To that end, if we contract the equation $\Psi_{abcdefhi} = 0$, with $X_1^{\ a}X_2^{\ c}X_3^{\ e}X_4^{\ i}X_4^{\ b}X_4^{\ d}X_1^{\ f}X_1^{\ h}$, we see that the result is $\alpha_1 = 0$. Similarly, multiplying $\Psi_{abcdefh} = 0$, with $X_1^{\ a}X_2^{\ c}X_3^{\ e}X_4^{\ i}X_4^{\ b}X_4^{\ f}X_1^{\ h}X_1^{\ d}$ and $X_1^{\ a}X_2^{\ c}X_3^{\ e}X_4^{\ i}X_4^{\ b}X_4^{\ h}X_1^{\ d}X_1^{\ f}$, gives us $\alpha_2 = \alpha_3 = 0$. Using a similar strategy it is possible to show that $\Psi_{abcdefhi} = 0$, implies that all of the $\alpha$'s must vanish. Hence the set of tensors generating $V_1$ is indeed a linearly independent set.

The next step in our construction of $\Psi_{abcdefhi}$, is to demand that the expression for $\Psi_{abcdefhi}$ given in EqC.17, satisfies the symmetries presented in Eq.C.15. When that is done, we need to determine what happens to the $\alpha$ coefficients in Eq.C.17 when we require that $\Psi_{abc(defh)i} = 0$. It turns out that once this demand is satisfied there is only one independent $\alpha$ left, and the expression for $\Psi_{abcdefhi}$ is equal to the one presented in Lemma 6. There is no need to check the equation $\Psi_{a(b|cd|efh)i} = 0$ is satisfied, since it turns out to be redundant in this case .

You have probably realized that I am getting ahead of myself here. All we have done so far is build the general elements of $V_1$ and $V_2$. What about $V_0$? I had hoped that you forgot about that, since determining what is in $V_0$ is quite onerous, and it turns out that there is nothing in there! But that requires proof.

The form of an ansatz element, $\Psi_{abcdefhi}$ , in $V_0$, has 105 terms in it to start. The general form of such a tensor, before we impose symmetry constraints, is given by

$$\Psi_{abcdefhi} =$$

$\alpha_1(ab)(cd)(ef)(hi)+ \ \alpha_2(ab)(cd)(eh)(if)+ \alpha_3(ab)(cd)(ei)(fh)+ \alpha_4(ab)(ce)(fh)(id)+$

$+ \alpha_5(ab)(ce)(fi)(dh)+ \ \alpha_6(ab)(ce)(fd)(hi)+ \alpha_7(ab)(cf)(hi)(de)+ \alpha_8(ab)(cf)(hd)(ei)+$

$+\alpha_9(ab)(cf)(he)(id)+\alpha_{10}(ab)(ch)(id)(ef)+\alpha_{11}(ab)(ch)(ie)(fd)+\alpha_{12}(ab)(ch)(if)(de)+$

$+\alpha_{13}(ab)(ci)(de)(fh)+\alpha_{14}(ab)(ci)(df)(he)+\alpha_{15}(ab)(ci)(dh)(ef)+\alpha_{16}(ac)(de)(fh)(ib)+$

$+\alpha_{17}(ac)(de)(fi)(bh)+\alpha_{18}(ac)(de)(fb)(hi)+\alpha_{19}(ac)(df)(hi)(be)+\alpha_{20}(ac)(df)(hb)(ei)+$

$+\alpha_{21}(ac)(df)(he)(ib)+\alpha_{22}(ac)(dh)(ib)(ef)+\alpha_{23}(ac)(dh)(ie)(fb)+\alpha_{24}(ac)(dh)(if)(be)+$

$+\alpha_{25}(ac)(di)(be)(fh)+\alpha_{26}(ac)(di)(bf)(he)+\alpha_{27}(ac)(di)(bh)(ef)+\alpha_{28}(ac)(db)(ef)(hi)+$

$+\alpha_{29}(ac)(db)(eh)(if)+\alpha_{30}(ac)(db)(ei)(fh)+\alpha_{31}(ad)(ef)(hi)(bc)+\alpha_{32}(ad)(ef)(hb)(ci)+$

$+\alpha_{33}(ad)(ef)(hc)(ib)+\alpha_{34}(ad)(eh)(ib)(cf)+\alpha_{35}(ad)(eh)(ic)(fb)+\alpha_{36}(ad)(eh)(if)(bc)+$

$+\alpha_{37}(ad)(ei)(bc)(fh)+\alpha_{38}(ad)(ei)(bf)(hc)+\alpha_{39}(ad)(ei)(bh)(cf)+\alpha_{40}(ad)(eb)(cf)(hi)+$

$+\alpha_{41}(ad)(eb)(ch)(if)+\alpha_{42}(ad)(eb)(ci)(fh)+\alpha_{43}(ad)(ec)(fh)(ib)+\alpha_{44}(ad)(ec)(fi)(bh)+$

$+\alpha_{45}(ad)(ec)(fb)(hi)+\alpha_{46}(ae)(fh)(ib)(cd)+\alpha_{47}(ae)(fh)(ic)(db)+\alpha_{48}(ae)(fh)(id)(bc)+$

$+\alpha_{49}(ae)(fi)(bc)(dh)+\alpha_{50}(ae)(fi)(bd)(hc)+\alpha_{51}(ae)(fi)(bh)(cd)+\alpha_{52}(ae)(fb)(cd)(hi)+$

$+\alpha_{53}(ae)(fb)(ch)(id)+\alpha_{54}(ae)(fb)(ci)(dh)+\alpha_{55}(ae)(fc)(dh)(ib)+\alpha_{56}(ae)(fc)(di)(bh)+$

$+\alpha_{57}(ae)(fc)(db)(hi)+\alpha_{58}(ae)(fd)(hi)(bc)+\alpha_{59}(ae)(fd)(hb)(ci)+\alpha_{60}(ae)(fd)(hc)(ib)+$

$+\alpha_{61}(af)(hi)(bc)(de)+\alpha_{62}(af)(hi)(bd)(ec)+\alpha_{63}(af)(hi)(be)(cd)+\alpha_{64}(af)(hb)(cd)(ei)+$

$+\alpha_{65}(af)(hb)(ce)(id)+\alpha_{66}(af)(hb)(ci)(de)+\alpha_{67}(af)(hc)(de)(ib)+\alpha_{68}(af)(hc)(di)(be)+$

$+\alpha_{69}(af)(hc)(db)(ei)+\alpha_{70}(af)(hd)(ei)(bc)+\alpha_{71}(af)(hd)(eb)(ci)+\alpha_{72}(af)(hd)(ec)(ib)+$

$+\alpha_{73}(af)(he)(ib)(cd)+\alpha_{74}(af)(he)(ic)(db)+\alpha_{75}(af)(he)(id)(bc)+\alpha_{76}(ah)(ib)(cd)(ef)+$

$+\alpha_{77}(ah)(ib)(ce)(fd)+\alpha_{78}(ah)(ib)(cf)(de)+\alpha_{79}(ah)(ic)(de)(fb)+\alpha_{80}(ah)(ic)(df)(be)+$

$+\alpha_{82}(ah)(ic)(db)(ef)+\alpha_{82}(ah)(id)(ef)(bc)+\alpha_{83}(ah)(id)(eb)(cf)+\alpha_{84}(ah)(id)(ec)(fb)+$

$+\alpha_{85}(ah)(ie)(fb)(cd)+\alpha_{86}(ah)(ie)(fc)(db)+\alpha_{87}(ah)(ie)(fd)(bc)+\alpha_{88}(ah)(if)(bc)(de)+$

$+\alpha_{89}(ah)(if)(bd)(ec)+\alpha_{90}(ah)(if)(be)(cd)+\alpha_{91}(ai)(bc)(de)(fh)+\alpha_{92}(ai)(bc)(df)(he)+$

$+\alpha_{93}(ai)(bc)(dh)(ef)+\alpha_{94}(ai)(bd)(ef)(hc)+\alpha_{95}(ai)(bd)(eh)(cf)+\alpha_{96}(ai)(bd)(ec)(fh)+$

$+\alpha_{97}(ai)(be)(fh)(cd)+\alpha_{98}(ai)(be)(fc)(dh)+\alpha_{99}(ai)(be)(fd)(hc)+\alpha_{100}(ai)(bf)(hc)(de)+$

$+\alpha_{101}(ai)(bf)(hd)(ec)+\alpha_{102}(ai)(bf)(he)(cd)+\alpha_{103}(ai)(bh)(cd)(ef)+\alpha_{104}(ai)(bh)(ce)(fd)+$

$+\alpha_{105}(ai)(bh)(cf)(de)$ , $\hspace{3cm}$ Eq.C.18

where the α's are scalar functions of φ.

Before proceeding, you should examine how I presented the above terms in $\Psi_{abcdefhi}$. I have presented them in essentially the same way that one would count how many terms are in $\Psi_{abcdefhi}$. That can be done as follows. The index "a" must appear somewhere, so we shall place it in the first of the four commuting g's in Ψ. Now there are seven indices left to place. Choose one and place it next to "a" in the first pair. There are six indices left to choose. Choose any one of them, and place it in the second g. Now there are five indices left that can be placed next to this index. Choose one, and there are four indices remaining. Take any one of those four and place it in the third g in Ψ. Lastly there are three indices left to place in the third g. Choose one, and then what goes in the fourth g is fixed. So there are 7x5x3 = 105, choices to be made, which accounts for the 105 terms in our initial expression for Ψ. (Note that in a similar way, we can show that any (0,2k) tensor built from k g's has 3x5x...x(2k‒1) terms in the initial expression for it.)

The next thing we have to check is that the terms which generate Ψ form a linearly independent set. This is an elementary task and its proof will be omitted. Now the really tedious part begins. You have to determine the form of $\alpha_1, . . . , \alpha_{105}$ so that Eqs.C.15 and C.16 are satisfied. Lengthy calculations show that the only way that these condions are met is if all of the α's vanish. Hence we have shown that $V_0 = \{0\}$.

Thus our construction of $\Psi_{abcdef}$ is now complete.

The last thing we need to do to complete the proof of Lemma 6 is to construct $\Phi^{abcd}$. You can do this quite simply by using Weyl's theory.

In Lemma 13 we need to construct $\Psi^{abcdefhi}$ which is a tensorial concomitant of $g_{ab}$ and $\varphi$, and possesses symmetries slightly different then those given in Eqs.C.15 and C.16, (*see*, Lemma 12). These symmetries preclude the appearance of one or two $\varepsilon$'s in $\Psi$, and so it must be built from the product of four $g$'s. One can use Eq.C.18 to build $\Psi^{abcdefhi}$. Once again I omit the copious computational details.

Eqs.2.48 and 2.49 require us to evaluate $\Psi_1^{abcdef}R_{cabd}\,\varphi_{ef}$ and $\Psi_2^{abcdef}R_{cabd}\varphi_e\varphi_f$. This can be done without actually determining the explicit functional form of either $\Psi_1$ or $\Psi_2$, in the following way. From Weyl we know that $\Psi_1$ and $\Psi_2$ must be built from the sum of products of one $g$ and one $\varepsilon$, or three $g$'s, with coefficients which are scalar functions of $\varphi$. There is no way to have an $\varepsilon$ in either one of the quantities in question, since it would lead to zero when summed into either $R_{cabd}\varphi_{ef}$ or $R_{cabd}\varphi_e\varphi_f$. Thus $\Psi_1$ and $\Psi_2$ must be built from the product of three $g$'s. When any such combination is summed into the $R_{cabd}\varphi_{ef}$, or $R_{cabd}\varphi_e\varphi_f$ terms, the result is either zero, or the terms presented on the right-hand sides of Eqs.2.48 and 2.49. This explains how I arrived at those equations.

Of the two approaches to constructing tensorial concomitants of $g_{ab}$ and $\varphi$,

which I have described in this appendix, I believe that, in general, the algebraic approach is the simplest, although both require some effort. Nevertheless, there are occasions when it is convenient to combine the two approaches. *E.g.,* say you need to construct a (0,10) tensorial concomitant of $g_{ab}$ and $\varphi$. Then the general term in $V_0$ would have 945 terms involving the product of five g's. What is worse, is that these terms would not form a linearly independent set of (0,10) tensors. In this case you might appeal to the differential approach, which hopefully would reduce your task to building (0,8) tensorial concomitants, which you could construct with the algebraic approach, as above.

In [1] I was confronted with the task of building tensorial concomitants of $g_{ab}$, $\varphi$ and $\varphi_{,a}$. I employed the differential approach to tackle that problem. I do not know if the algebraists have proved that such concomitants must be built from all possible products of $g_{ab}$, $\varphi_{,a}$ and $e_{abcd}$ with coefficients which are scalar function of $\varphi$ and $\rho$. Let's hope so.